\documentclass[superscriptaddress,nobibnotes,amsmath,amssymb,notitlepage,twocolumn,prl,longbibliography,floatfix]{revtex4-1}

\usepackage{bm,mathptmx,braket}

\usepackage{graphicx}
\usepackage[caption=false]{subfig}

\usepackage[usenames,dvipsnames]{xcolor}
\usepackage{soul}
\usepackage{amsmath}
\usepackage{amsfonts}
\usepackage{amssymb}
\usepackage[colorlinks=true,citecolor=blue,linkcolor=magenta,filecolor=magenta]{hyperref}
\usepackage[capitalize]{cleveref}
\usepackage[makeroom]{cancel}		% crossed terms in tables
\usepackage{multirow}				% merge cells vertically in tables
\usepackage[normalem]{ulem}         % strike-through text
\usepackage{array}
\usepackage{mathrsfs}
\usepackage{mathtools}
\usepackage[version=4]{mhchem}      % for chemical formulae
\usepackage{IEEEtrantools}          % for better equation arrays
\usepackage{dcolumn}                % for cell alignment at decimal point
\usepackage{dsfont}                 % for double stroke font, e.g. the identity matrix
\usepackage[shortlabels]{enumitem}  % for enumerations
\usepackage{verbatim}
\usepackage{csquotes}
\usepackage{pifont}

\usepackage{mathdots}

\newcommand{\beq}[1]{\begin{equation}\label{#1}}
\newcommand{\eep}{\;.\end{equation}}
\newcommand{\eec}{\;,\end{equation}}
\newcommand{\eeq}{\end{equation}}

\newcommand{\lb}{\left(}
\newcommand{\rb}{\right)}

\newcommand{\G}{\Gamma}

\newcommand{\sect}[1]{\vspace{0.3em}{\it #1.}---}

\DeclareMathAlphabet{\mathcal}{OMS}{cmsy}{m}{n} % Changes font for mathcal but leaves the rest of the math fonts in Times.

\newcommand{\F}{\mathcal{F}}    % Free energy
\renewcommand{\vec}[1]{{\bf #1}}

\newcommand{\kv}{\vec{k}}
\newcommand{\rv}{\vec{r}}

\newcommand{\q}{\vec{q}}
\newcommand{\Q}{\vec{Q}}

\def\app#1#2{%
  \mathrel{%
    \setbox0=\hbox{$#1\sim$}%
    \setbox2=\hbox{%
      \rlap{\hbox{$#1\propto$}}%
      \lower1.1\ht0\box0%
    }%
    \raise0.25\ht2\box2%
  }%
}
\def\approxprop{\mathpalette\app\relax}

\begin{document}

\title{Geometric Ginzburg-Landau theory of charge ordering and commensurability}

%%%% AFFILIATIONS %%%s
\newcommand{\UoM}{{Department of Physics and Astronomy, University of Manchester, Oxford Road, Manchester M13 9PL, United Kingdom}}

\newcommand{\TCM}{{Theory of Condensed Matter Group, Cavendish Laboratory, University of Cambridge, J.\,J.\,Thomson Avenue, Cambridge CB3 0HE, United Kingdom}}

%%% AUTHORS %%%

\author{Aneesh Agarwal}
\email{aa2223@cam.ac.uk}
\affiliation{\TCM}

\author{Rutvij Gholap}
\affiliation{\UoM}

\author{Mohammad Saeed Bahramy}
\affiliation{\UoM}

\author{Robert-Jan Slager}
\email{robert-jan.slager@manchester.ac.uk}
\affiliation{\UoM}
\affiliation{\TCM}

\date{\today}

\begin{abstract}
The concept of quantum geometry~\cite{provost1980riemannian,ma2010nonabelian,bouhon2023quantum} has recently led to reinvigorated insights in a wide range of fields including physical responses, superconductivity, and optical transitions, with effects most pronounced in systems with nearly flat dispersion~\cite{Yu2025review,Peotta2015,arbeitman2022superfluid,marzari1997spread,resta2011geometric,ParameswaranRoySondhi2013fractional,roy2014fractional,Fuijimoto2025vortexabilty,Ahn2021_riemannian}. Here, we show that it plays an essential role in charge density wave formation (CDW) -- an important physical phenomenon that is responsible for driving various sharp changes in material transport properties including metal-insulator transitions~\cite{gruner1988cdw,Di_Salvo1979cdw,imada1998MIT,FriendJerome1979pldcdw}. We derive an effective Ginzburg-Landau theory including uncharted contributions and, as a highlight, discover a general criterion for both CDW formation and commensurability transitions where underlying electron-phonon interactions manifest purely as electronic quantum geometric enhancements/suppressions. 
We benchmark our framework in a class of transition-metal dichalcogenides and resolve a longstanding puzzle where well-established purely kinetic CDW criteria fail in describing the correct ordering wavevector \cite{Peierls1955,zhu2015_cdwclassification,Johannes_nesting_2008,flicker_nbse2_2016}. 
Besides rendering robust criteria and fundamental insights that are immediately relevant to several experimental charge ordering systems~\cite{moncton_nbse2_1977,Moncton_nbse2_1975,Liu_tas2_expt}, our theory can also be applied directly to other phonon-mediated phases such as superconductivity, and can be used as an important tool to explore the interplay between various such states. More generally, our framework provides a recipe for investigating the role of quantum geometry in phase transitions.
\end{abstract}

\maketitle
\sect{Introduction}
Topology has provided a revolutionizing view on the characterization of quantum matter in the past decades. While topology, especially in nearly free fermion systems~\cite{Hasan2010rmp,Qi2011rmp,Kruthoff2017,Po2017,bouhon2020geometric,Bradlyn2017}, has increasingly been understood, quantum geometry~\cite{provost1980riemannian,ma2010nonabelian,bouhon2023quantum,Yu2025review} has recently been galvanizing novel insights. The surge of interest in flatband systems that naturally provide for a playground in which interaction effects can enjoy synergies with geometrical and topological effects have led to several developments. In the absence of dispersive effects, the physics becomes inherently geometrically motivated, and effects on superconductivity and superfluid stiffness~\cite{liang.superfluid.stiffness,Peotta2015,arbeitman2022superfluid}, wavefunction spreading~\cite{marzari1997spread,resta2011geometric}, fractionalization~\cite{ParameswaranRoySondhi2013fractional,roy2014fractional,Fuijimoto2025vortexabilty}, as well as optical responses~\cite{Ahn2021_riemannian, Jankowski2024,KrishnaKumar2025photocurrent} are actively being explored through this geometrical lens.

These developments motivate a reinvigoration of well established phenomena as well as longer standing questions in such contexts. A key example in this regard concerns charge density wave formation in systems with a small bandwidth as compared to the band gap. Charge density waves (CDWs) \cite{gruner1988cdw,Gruner2019book,Di_Salvo1979cdw} represent a main field of research as one of the driving mechanisms behind metal-insulator transitions and other sharp changes in transport properties~\cite{FriendJerome1979pldcdw,imada1998MIT,FazekasTosatti19791Ttas2}, and due to the evolution of scanning tunneling microscopy (amongst others) \cite{wang1990stm,arguello2014stm,soumyanarayan2013nbse2stm}, are observed on a daily basis. Although the foundations for such phenomena can be traced to Fermi surface nesting that underlies the Peierls instability \cite{Peierls1955} and the Kohn anomaly \cite{Kohn1959}, experiments have shown that charge ordering can also occur in systems where such criteria are not met \cite{mcmillan1977tase2,zhu2015_cdwclassification,johannes_nbse2_2006, Johannes_nesting_2008,flicker_nbse2_2016,Moncton_nbse2_1975,moncton_nbse2_1977,wilson1974_expt}. What is more surprising perhaps is that there seem to be no hints in the electronic spectra regarding either the ordering itself or a preference of the system to a specific soft phonon wavevector \cite{zhu2015_cdwclassification,flicker_nbse2_2016}. This naturally begs the question that geometry might be the driving factor, especially since these inconsistencies are most famously reported in relatively flat systems such as $2\text{H}$ polymorphs of transition metal dichalcogenides (TMDs). While some recent work~\cite{kivelson, balents} addressed that nesting conditions can indeed have a geometric component, their direct relation to classes of electron-phonon coupled Hamiltonians, and consequently real material modeling, has remained more elusive. 
TMDs in particular stand out because the CDW formation is known to not uniformly match these conditions in certain cases.

\begin{figure*}[t]
    \centering
    \includegraphics[width=0.95\linewidth]{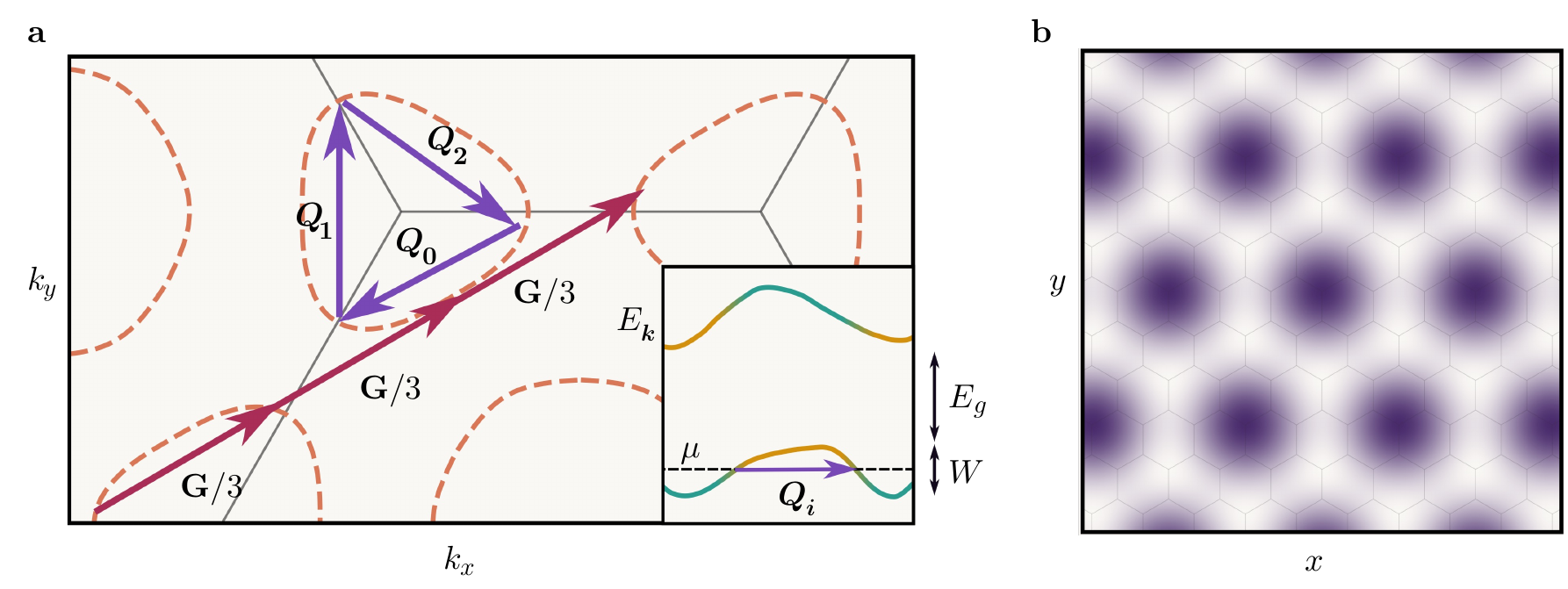}
    \caption{\textbf{Geometric mechanism for charge-density-wave formation in flat-band systems.} \textbf{a} Schematic illustration of the geometric origin of charge ordering. In flat and quasi-flat band systems, charge-density-wave instabilities are enhanced when Bloch states at momenta separated by an ordering wavevector $\Q_i$ undergo a strong change in flavor character, as illustrated by the color exchange in the band-structure inset. In hexagonal systems, three symmetry-related wavevectors, $\Q_0$, $\Q_1$, and $\Q_2$, further enable a higher-order triangular scattering process (purple) that contributes to wavevector selection. At commensurate wavevectors $\Q_i=\G/3$, interference between triangular and Umklapp (magenta) scattering processes governs the commensurability transition. \textbf{b} Corresponding real-space charge-density-wave modulation resulting from the selected ordering wavevector.}
    %\caption{\textbf{Geometric and higher-order contributions to charge ordering.} We find that charge ordering in flat band systems can be driven by large changes in the electronic flavor character (portrayed by the colours in the bandstructure) near the chemical potential. These are strongest when the Bloch states at $\kv$ and $\kv+\Q_i$ completely exchange their character as shown in the inset. Additionally, in (usually) hexagonal systems, the ordering wavevector can be selected by a higher order scattering mechanism involving three states on the Fermi surface as shown. At commensurate $\Q_i$, the interference between the triangular (purple) and umklapp (magenta) scattering mechanisms determines a commensurability transition.}
    \label{fig:fig1}
\end{figure*}

Here, we develop a definitive geometric theory for phases driven by electron-phonon interactions. Beginning with a generic electron-phonon Hamiltonian \cite{mcmillan1968lambda,varma}, where the electronic quantum geometry manifests through the interaction form factors \cite{yu2024epc.geo}, we derive a geometric Ginzburg-Landau (GL) theory for CDW formation. Particularly, we find both dominant higher-order terms and geometric modifications to terms at all orders in the GL functional that drive CDW formation in materials long-thought to have no conclusive indicators of ordering in their electronic Hamiltonians. Since these materials often exhibit largely flat Lindhard susceptibilities (due to low dispersion), ordering finds its origins in large changes of the electronic flavor character (and hence, a large interband velocity matrix element) near the chemical potential. Specifically, we note that a system is most likely to condense into an ordered state at a wavevector $\q$ where Bloch states at $\q$-separated momenta on the Fermi surface exchange their flavor character or equivalently, rotate into each other (see Figure \ref{fig:fig1}). Further, we find that ordering can be enhanced by the presence of a third state on the Fermi surface connected via a symmetrically-equivalent $\q$-vector, something that is highly likely in hexagonal systems which often have roughly triangular Fermi pockets (see Figure \ref{fig:fig1}). 

As a highlight, we also discover that at commensurate ordering wavevectors $=1/3 \, \vec{G}$, umklapp scattering can interfere with the aforementioned triangular scattering process (see Figure \ref{fig:fig1}), and the relative phase determines whether a commensurability condition can be expected. Notably, the phases of such scattering processes are purely geometric, thus suggesting that a commensurability transition is not a feature of the energetics at all. We benchmark our framework through ab initio calculations in several TMDs where the origin of the ordering wavevector has long been debated and find geometry-driven ordering peaks at experimentally-observed $\q$ along with associated commensurability transitions (where present). Further, where ordering peaks have similar magnitude, we predict multiple co-existing CDW phases in line with STM observations \cite{soumyanarayan2013nbse2stm}. 

%This allows us to solve longstanding puzzles. We in particular benchmark our theory in the context of TMDs and find a generic thoery to address CDWs.

\sect{Theory setting}
The effects of electronic band geometry on CDW formation can be traced to the form of the electron-phonon interaction. Generically, the interaction term in the electron-phonon Hamiltonian $H_{\text{el-ph}}$ can be written in second-quantized form as \cite{mcmillan1968lambda}

\begin{equation}
    H_{\text{el-ph}} = \sum_{\kv,\q,m,n,\nu} \gamma_{\q\nu} \ \vec{e}_{\q\nu}\cdot \bm{\lambda}_{\kv,\kv-\q}^{mn} \lb \hat{b}_{\q\nu} + \hat{b}^{\dagger}_{-\q\nu} \rb \hat{c}^{\dagger}_{m,\kv} \hat{c}_{n,\kv-\q}, 
\end{equation}

\noindent where $\vec{e}_{\q\nu}$ is the phonon polarization at wavevector $\q$ and branch $\nu$, $\gamma_{\q\nu}$ is some term dependent only on the phonon spectrum (usually described as scaling with the phonon frequency $\Omega_{\q\nu}$ as $\Omega_{\q\nu}^{-1/2}$), and $\hat{b}_{\q\nu}$ and $\hat{c}_{m,\kv}$ are the phonon and electron annihilation operators at their respective wavevectors and branch/band quantum numbers. Ref. \cite{varma} found the electron band geometry factor to be $\bm{\lambda}_{\kv,\kv'}^{mn} \propto -i\left[U^{\dagger} (\kv) \lb \nabla_\kv H_0(\kv) - \nabla_{\kv'} H_0(\kv') \rb U(\kv')\right]_{mn}$, where $H_0(\kv)$ is the bare electronic Hamiltonian at wavevector $\kv$ which is diagonalized in the Bloch basis by unitary matrices $U_{\alpha n}(\kv)$. 

Integrating out the phononic degrees of freedom, the effective electronic Hamiltonian is found to contain an interaction term given by 

\begin{equation} \label{eqn:H_int}
    H_{\text{int}} = -\sum_{\q} \hat{\bm{\rho}}_{\q} \lb\sum_\nu \textit{D}_{0,\q\nu} |\gamma_{\q\nu}|^2 \vec{e}_{\q\nu} \otimes \vec{e}_{\q\nu}^* \rb \hat{\bm{\rho}}_{-\q},
\end{equation}

\noindent where $\hat{\bm{\rho}}_{\q} = \sum_{m,n,\kv} \bm{\lambda}_{\kv,\kv-\q}^{mn}\hat{c}^{\dagger}_{m,\kv}\hat{c}_{n,\kv-\q}$ is an appropriately weighted density operator and $\textit{D}_{0,\q\nu}$ is the bare phonon propagator. The interaction vertex $\frac{1}{4}g_\q = \sum_\nu \textit{D}_{0,\q\nu} |\gamma_{\q\nu}|^2 \vec{e}_{\q\nu} \otimes \vec{e}_{\q\nu}^*$ consists of the summation of the weighted phonon propagator across all phonon branches and is therefore, expected to be a full rank tensor with a well-defined inverse. For the purposes of deriving a CDW nesting-like criterion dependent on only the electronic Hamiltonian, $g_\q$ will be assumed to be fairly flat over the $\q$ states of interest. While this is not expected to be the case generally, such an assumption allows for an isolation of the cause of any experimentally-observed instability to be either the bare electronic Hamiltonian (as we study here) or some strong phononic driving factors. 

Decoupling in the density channel via a Hubbard-Stratonovich field $\bm{\phi}_\q$ (equal to $\sum_{m,n,\kv}\langle g_\q \bm{\lambda}_{\kv,\kv-\q}^{mn}\hat{c}^{\dagger}_{m,\kv}\hat{c}_{n,\kv-\q} \rangle$ at mean field level) and expanding near $\bm{\phi}_\q=0$, a Ginzburg-Landau functional can be obtained as (see Methods) 

\begin{equation} \label{eqn:GL}
    \mathcal{F} =  \sum_\q \bm{\phi}_\q \, g_\q^{-1} \bm{\phi}_{-\q} + \sum_{n,\{\q_i\}} f^{(n)}_{ij...l} \{\q_i\} \bm{\phi}_{\q_i}^i\bm{\phi}_{\q_j}^j...\bm{\phi}_{\q_l}^l,
\end{equation}

\noindent where the $\{\q_i\}$ satisfy $\sum_i^n \q_i = \vec{G}$ (for reciprocal lattice vectors $\vec{G}$), and each coefficient $f^{(n)}$ in the expansion is an $n$-index contravariant tensor derived in Methods. Particularly, Equation \eqref{eqn:GL} implies that the standard Lindhard susceptibility (ordinarily appearing in the coefficient of the quadratic term) is now modified to a generically rank-2 tensor given by 

\begin{equation} \label{eqn:chi}
    \chi (\q) = \frac{1}{2}\sum_{m,n,\kv} \frac{n_F(\xi_{n,\kv}) - n_F(\xi_{m,\kv+\q})}{\xi_{m,\kv+\q}-\xi_{n,\kv}}\bm\lambda^{mn}_{\kv,\kv+\q}\otimes\bm\lambda^{mn\ *}_{\kv,\kv+\q},
\end{equation}

\noindent where $n_F(\xi_{n,\kv}=E_n(\kv)-\mu)$ is the Fermi-Dirac distribution function representing the occupancy of the band $n$ with energy $E_n(\kv)$ at chemical potential $\mu$. 

For a CDW at a single wavevector, mean-field theory would then suggest that the system is susceptible to charge ordering at wavevector $\q^*$, where the largest eigenvalue of $\chi$ is maximized. While systems with perfect nesting would still be dominated by the divergence of the energetic factor, suppression due to the geometric $\bm\lambda$ factors assumes a significant role in imperfectly nested and flatband systems. 

Writing the Hamiltonian in terms of single-band projectors $P^n_{\alpha\beta}(\kv) = U_{\alpha n}(\kv)U^{\dagger}_{n\beta}(\kv)$ as $H_0(\kv)=\sum_n E_n(\kv)P^n(\kv)$), the velocity operator in $\bm\lambda^{mn}_{\kv\kv'}$ can be separated into a band velocity term $\bm v_1(\kv)$ and a purely geometric term $\bm v_2(\kv)$ since $\partial_\kv H_0(\kv) = \sum_n (\partial_\kv E_n(\kv)P^n(\kv)+E_n(\kv)\partial_\kv P^n(\kv))$. In perfectly flatband systems where $\partial_\kv E_n(\kv)$ and hence, $\bm v_1(\kv)$ is vanishing everywhere in the BZ, $\bm\lambda^{mn}_{\kv\kv'}$ for the flat bands can be rewritten approximately as $\bm\lambda^{mn}_{\kv\kv'} \approx -iE_g \left[U^\dagger(\kv) \lb \partial_\kv \bar{P}_\kv - \partial_{\kv'} \bar{P}_{\kv'}  \rb U(\kv')\right]_{mn}$, where $E_g$ is the bandgap between the flat bands and the higher bands (see Methods). $P_\kv$ is the projector onto the flatband subspace $S_\kv$ at wavevector $\kv$, while $\bar{P}_{\kv} = 1-P_\kv$ projects onto its complement $\bar{S}_\kv$. The quadratic susceptibility in Equation \eqref{eqn:chi} is then (see Methods)

\begin{equation} \label{eqn:chi_geo}
    \chi(\q) \approxprop E_g^2 \sum_\kv \text{Tr} \left[P_\kv \bar{\mathcal{Q}}_{\kv+\q} + (\kv \leftrightarrow \kv+\q)^\dagger \right],
\end{equation}

\noindent where $\bar{\mathcal{Q}}_{\kv,\mu\nu} = \bar{P}_\kv (\partial_{k_\mu} P_\kv \partial_{k_\nu} P_\kv) \bar{P}_\kv$ is the non-Abelian quantum geometric tensor (QGT) in the subspace $\bar{S}$ \cite{provost1980riemannian,ma2010nonabelian}, and the trace is over band indices only. Importantly, the susceptibility of the system to charge ordering at wavevector $\q$ is determined solely by the overlap between the QGT and the flatband projector at $\q$-separated quasimomenta. 

In hexagonal lattices, if the $C_3$ symmetry remains unbroken, ordering must occur at three wavevectors related by $2\pi/3$ rotations, henceforth called triple-Q solutions. Further, if these solutions are parameterized as $\bm\phi_\Q = \Delta_\Q \hat{\bm n}_\Q e^{i\alpha_\Q}$ with $\Delta_\Q \in \mathbb{R}$ and $\hat{\bm n}_\Q \in S^1$, $C_3$ enforces the order parameter at $\Q_0$ and rotated counterparts $\Q_n=R_{2\pi n/3} \Q_0$ to be related by $\Delta_{\Q_n} = \Delta_{\Q_0}$ and $\hat{\bm n}_{\Q_n} = R_{2\pi n/3}\hat{\bm n}_{\Q_0}$ (see Methods for an irreducible representation analysis where we find this to be the most general symmetry-allowed form). Explicitly looking for such solutions, the free energy at an ordering wavevector $\Q \in \{\Q_0,\Q_1,\Q_2 \}$ can be written to fourth order as 

\begin{equation} \label{eqn:F_tripleQ}
    \F = \left(\frac{3}{\text{Tr} \, g}+\Pi\right) \Delta^2+\frac{1}{2}\left(\gamma \Delta^3 e^{i\sum_{n=0}^2 \alpha_{n}}+\text{c.c.}\right)+\beta\Delta^4, 
\end{equation}

\noindent where the coefficients $\Pi$, $\gamma$, and $\beta$ involve contractions of the coefficients $f^{(n)}_{ij...l} \{\Q_n\}$ in Equation \eqref{eqn:GL} with $\hat{\bm n}_{\Q_n}$ and can be derived from the loop diagrams shown in Figure \ref{fig:loops} (see Methods). Particularly, $\Pi_\Q$ is proportional to the smallest eigenvalue of the modified Lindhard susceptibility in Equation \eqref{eqn:chi}. The dependence of all quantities on $\Q$ has been suppressed for clarity. 

Importantly, we note that, owing to the three $\Q$ solutions satisfying (due to the presence of $C_3$ symmetry) $\sum_{n=0}^2 \Q_n = 0$, the cubic term provides a nonzero contribution and drives a (weakly) first-order transition into the CDW state. Additionally, the phase $\alpha_{\Q_n}$ only enters the functional through the cubic term and only as $\sum_{n=0}^2 \alpha_{\Q_n}$. Minimizing $\mathcal{F}$ implies $\sum_{n=0}^2 \alpha_{\Q_n} = -\text{arg}(\gamma_{\Q})+\pi$, leaving two phases undetermined revealing the two gapless sliding phason modes for an incommensurate CDW in 2 dimensions. 

Setting the phase as above, the transition into the ordered CDW phase occurs when 

\begin{equation} \label{eqn:transition}
    3/\text{Tr} \, g_\Q = \Theta_\Q :=-\Pi_\Q+|\gamma_\Q|^2/4\beta_\Q,
\end{equation}

\noindent where the order parameter $\Delta_\Q$ discontinuously changes from $0$ to $|\gamma_\Q|/2\beta_\Q$. The wavevector of the CDW corresponds to the $\Q$ for which Equation \eqref{eqn:transition} is satisfied first as temperature is reduced. For fairly flat $g_\Q$, the system is thus unstable to ordering at the wavevector $\Q^*$ that maximizes $\Theta_\Q$, henceforth referred to as the `CDW driver'. Note that in the absence of any cubic term, the driver reduces to $-\Pi_\Q$ ($\propto \chi_\Q$), and the standard CDW instability condition is recovered where a peak in the Lindhard susceptibility corresponds to charge ordering. We emphasize however that in the absence of strong $\Q$-dependence/singular behavior of the Lindhard susceptibility, the inclusion of the cubic term in the driver is essential in selecting $\Q^*$. Further, in systems with small band velocities, the $\Q$-dependence of all coefficients in $\Theta_\Q$ is strongly modified by the quantum geometry. Considering the full CDW driver then becomes essential for analyzing flat and quasi-flat band systems (as observed, for example, in the TMDs considered in this work).

\begin{figure*}[t]
    \centering
    \includegraphics[width=\linewidth]{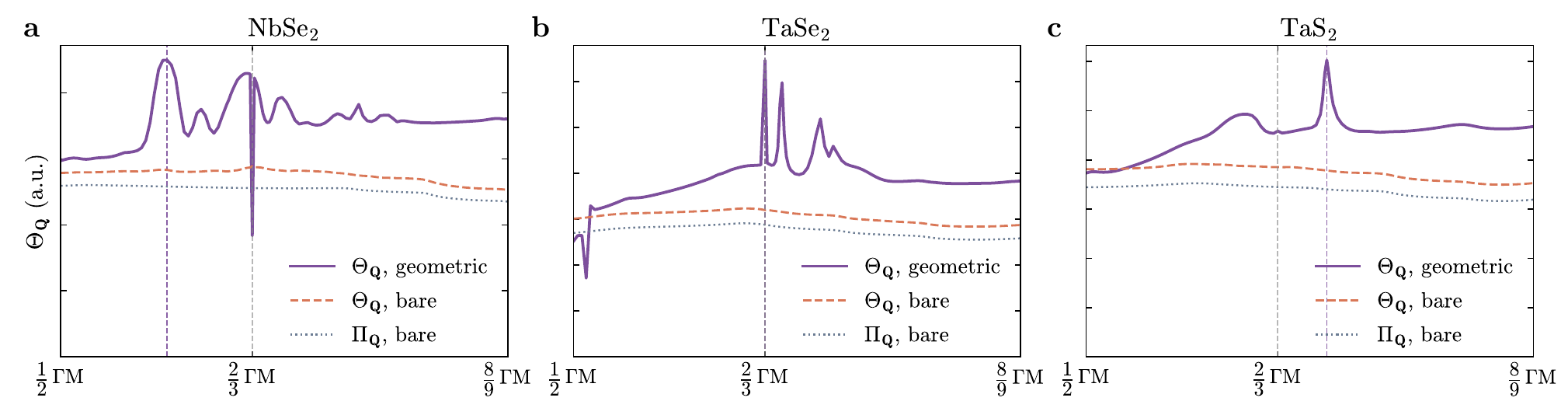}
    \caption{\textbf{CDW transition driver $\Theta_\Q$ as a function of $\Q$ on the $\Gamma \text{M}$ line for different materials.} $\Theta_\Q$ is plotted as a function of $\Q$ between $1/2 \, \Gamma \text{M}$ and $8/9 \, \Gamma \text{M}$ for \textbf{a} NbSe$_2$ ($T\approx 23 \, \text{K}$), \textbf{b} TaSe$_2$ ($T\approx 35 \, \text{K}$), and \textbf{c} TaS$_2$ ($T\approx 49 \, \text{K}$), and peaks near/at $2/3 \, \Gamma \text{M}$ are observed for all three materials. The $\Q$-domain is so chosen since the bare Lindhard susceptibility (dotted) exhibits a broad peak here and is inconclusive in determining the CDW wavevector. Inclusion of higher-order terms in just their energetic form (orange dashed line) is still inconclusive. The geometric enhancements studied here (purple solid line) then become the dominant mechanism behind CDW formation. Importantly, note that an incommensurate CDW is predicted for both NbSe$_2$ ($\Q$ just below $2/3 \, \Gamma \text{M}$ as well as an additional coexisting $\Q$ phase near $0.59 \, \Gamma \text{M}$) and TaS$_2$ ($\Q = 0.71 \, \Gamma \text{M}$), whereas a commensurate CDW is predicted for TaSe$_2$, similar to experiment. The peaks of the bare curves are scaled to the geometric curve for faithful comparison and displaced vertically for clarity.}
    \label{fig:fig2}
\end{figure*}

\sect{Universal commensurability condition}
As a highlight, we find that commensurate $\Q$ vectors contribute an additional diagram to the free energy summation in Equation \eqref{eqn:F_tripleQ}. Particularly, $\Q = 1/3 \vec{G}$, as commonly seen in experiment, contributes an additional diagram to the cubic term where equal-$\Q$ $\phi'$s multiply together since $\mathcal{G}_{0,\kv} = \mathcal{G}_{0,\kv+\vec{G}}$. This extra diagram depends on the individual phases of $\bm\phi_{\Q_n}$ through $e^{3i\alpha_{\Q_n}}$ (in contrast to just their sum) and thus, gaps the two phason modes as expected for a commensurate CDW. Additionally, residual $C_3$ symmetry forces the three phases to be equal ($\alpha_{\Q_n} = \alpha_{\Q_0}$) leading to $\gamma \rightarrow \gamma+\gamma'$ and $3 \alpha_\Q = {-\text{arg}(\gamma_\Q+\gamma'_\Q)+\pi}$. In such a case, the free energy for the commensurate CDW must be compared to the free energy of the optimal $\Q$ from Equation \eqref{eqn:F_tripleQ}, and a commensurability-incommensurability transition can be predicted. Explicitly, after fixing the phase as above, the cubic coefficient of $\Delta_\Q$ is given by (up to a constant factor) $|\gamma_\Q|$ for incommensurate CDW and by $|\gamma_\Q+\gamma'_\Q|$ for commensurate CDW. The commensurability transition is thus fixed by the relative phase between $\gamma_\Q$ and $\gamma'_\Q$.

\sect{Application of theory to TMDs}
We apply our framework to three TMD systems where the origin of CDW formation (in the absence of nesting) has been debated and largely attributed to phonon spectra. Particularly, the bare Lindhard susceptibility exhibits a very broad peak near $2/3 \, \Gamma \text{M}$ in these systems, thereby leading to conclusions that inhomogeneities in the phononic spectra would be responsible for selecting the CDW wavevector. Here, we show that the CDW transition, as well as a sharp dependence on $\Q$ to select the ordering wavevector, can actually be derived purely from the electronic Hamiltonian. 

In Figure \ref{fig:fig2}, we plot $\Theta_\Q$ in the region where the bare Lindhard susceptibility is flat and inconclusive, and show that the geometry-enhanced higher-order terms lead to sharp changes in the driver and are responsible for the CDW transition in these materials (see SI for plots of individual coefficients and variation with temperature). While the coefficients themselves and hence, $\Theta_\Q$ may get renormalized due to other interactions in the material, a sharp increase in the driver near $2/3 \, \Gamma \text{M}$ is strongly suggestive of a CDW transition at a wavevector near here. Interestingly, setting $\bm \lambda$ factors to be identity, the driver in just its energetic (bare) form (orange dashed line in Fig. \ref{fig:fig2}) is still insufficient (see SI for a detailed discussion), and consideration of the quantum geometry hiding in $\bm \lambda$ becomes essential for an accurate description of charge order in these materials. Further, an analysis of the temperature variation of $\Theta_\Q$ (see SI) reveals that up to the experimentally measured transition temperature \cite{FriendJerome1979pldcdw,Moncton_nbse2_1975,moncton_nbse2_1977, Liu_tas2_expt}, the quadratic term $\Pi_\Q$ in $\Theta_\Q$ remains largely constant. On the contrary, the geometry-enhanced higher-order terms vary significantly with temperature and are thus, primarily responsible for driving the CDW transition.

Concretely, writing the ordering wavevector as $\Q=(1+\delta)\cdot 2/3 \, \Gamma \text{M}$, we discover a peak just below $2/3 \, \Gamma \text{M}$ for NbSe$_2$ ($\delta=-0.01$), and a peak above $2/3 \, \Gamma \text{M}$ for TaS$_2$ ($\delta=0.06$); for TaSe$_2$, we find a commensurate peak at exactly $2/3 \, \Gamma \text{M}$, all of which closely match experimental results \cite{moncton_nbse2_1977,Moncton_nbse2_1975,Liu_tas2_expt}. For NbSe$_2$, two peaks of similar magnitude are found near $2/3 \, \Gamma \text{M}$ and $4/7 \, \Gamma \text{M}$ ($\delta = -0.01$ and $-0.11$ respectively), suggesting that different CDW orders may coexist, a prediction that is indeed corroborated in STM experiments \cite{soumyanarayan2013nbse2stm}. 

\begin{figure*}[t]
    \centering
    \includegraphics[width=\linewidth]{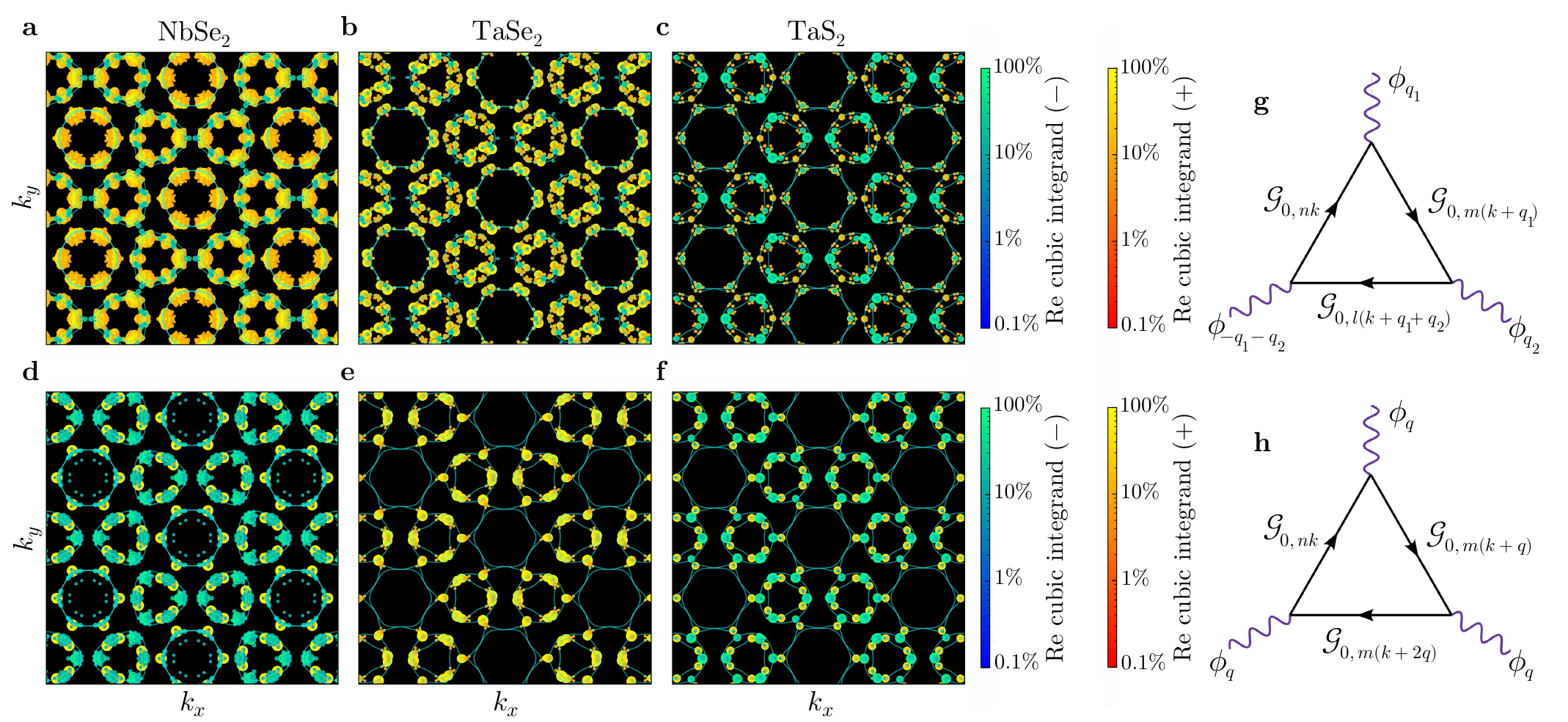}
    \caption{\textbf{Integrand for cubic coefficient at commensurate $Q=2/3\Gamma M$.} Normalized contributions to the coefficient of $\Delta^3$ in the Ginzburg-Landau free energy of different materials has been shown as a function of $\kv$ in the Brillouin Zone. The integrand consists of two terms: \textbf{a-c} triangle scattering processes where $\sum \Q_i=0$, and \textbf{d-f} umklapp scattering processes where $\sum \Q_i=\bm G$, a reciprocal lattice vector. Corresponding diagrams for the \textbf{g} triangle and \textbf{h} umklapp scattering are also sketched. The phase of the respective integrands is emphasized using different colors for positive and negative contributions. Interference between the triangle and umklapp scattering suggests incommensurate CDW in NbSe$_2$ and commensurate CDW in TaSe$_2$. For TaS$_2$, both constructive and destructive interference is present at different $\kv$, and thus, the extra diagram does not strongly affect the functional form for the CDW driver.}
    \label{fig:fig3}
\end{figure*}

The interference between the normal triangle scattering terms for the cubic coefficient and the additional umklapp scattering diagram for commensurate $\Q$ can be explicitly seen in the $\kv$-resolved integrand for the cubic coefficient (see Methods) in Figure \ref{fig:fig3}. We emphasize the phase difference between the triangle scattering terms (top row) and umklapp scattering terms (bottom row) by using different colors for positive and negative contributions to $\text{Re} \left(\gamma_\Q e^{i\sum_{n=0}^2 \alpha_{\Q_n}} \right)$. It is then easy to see that while NbSe$_2$ exhibits destructive interference (the triangle and umklapp scattering at the same $\kv$ add with opposite signs), the two terms in TaSe$_2$ add constructively. Since the triangle scattering can be expected to be smooth in $\Q$, destructive (constructive) interference with the umklapp scattering process in NbSe$_2$ (TaSe$_2$) suggests an incommensurate (commensurate) CDW, as seen in experiment. In TaS$_2$, the two terms add with differing relative signs at different $\kv$ points, and the extra diagram turns out to not affect the cubic coefficient significantly (as in the smooth behavior of the driver near $2/3 \, \Gamma \text{M}$ in Figure \ref{fig:fig2}). Such a weak dependence of the system on the commensurate umklapp diagram is likely to result in an incommensurate CDW as well (as corroborated by experiment).  

\sect{Discussions}
We discover two additional factors that contribute to charge ordering instabilities in imperfectly nested systems. In systems where the standard bare Lindhard susceptibility exhibits no sharp behavior as a function of ordering wavevector $\Q$, higher-order terms and geometric effects may combine to drive an instability. While the former modifies the transition condition by contributing an additive term on top of the quadratic susceptibility (see Equation \eqref{eqn:transition}), the latter modifies all terms in a significant $\Q$-dependent manner to drive the transition (see Figure \ref{fig:fig2}).

We emphasize that the dominant higher-order contribution to the transition comes from the cubic term which is only expected to be non-zero for triple-$\Q$ CDW states as in $C_3$-symmetric materials like the TMDs considered here (in contrast to single-$\Q$ or double-$\Q$ ordering in 1D or 2D orthorhombic/tetragonal lattices). Additionally, owing to the form of the cubic term (see Methods), we do not expect it to show any divergences (unless the system is perfectly nested $\xi_{\kv}=-\xi_{\kv+\q}$ and the quadratic term already diverges). Therefore, this contribution can only be expected to play a dominant role in imperfectly-nested hexagonal materials. In such materials, since the Fermi-surfaces themselves are $C_3$ symmetric, the cubic term connects three points on the Fermi surface only for a narrow band of $\Q$-vectors and is hence, the driving factor in selecting the ordering wavevector (see Figures \ref{fig:fig2} and \ref{fig:fig3}). 

Quantum geometric factors on the contrary may be dominant in enhancing or suppressing the susceptibility in all kinds of imperfectly nested materials. We reiterate that dependence on the electronic quantum geometry lives in the $\bm\lambda^{mn}_{\kv\kv'}$ factors, and a scaling analysis reveals when geometry can be expected to dictate the system's instability. $\bm v_1(\kv)$, which is dependent on the band velocity, can generally be expected to scale as $\sim Wa$, where $W$ is the bandwidth and $a$ is the lattice constant. Since $\partial_\kv H_0(\kv)$ scales as $\sim ta$, where $t$ is a measure of interatomic coupling, and $t\sim W$ usually, $\bm v_2(\kv)$ has a similar order of magnitude as $\bm v_1(\kv)$. However, in flatband systems where $W\rightarrow 0$, nontrivial geometry can cause $\bm v_2(\kv)\gg\bm v_1(\kv)$, and geometric effects will dominate. Additionally, the $W\rightarrow 0$ limit is also likely to result in flat generalized Lindhard susceptibilities appearing in the $f^{(n)}_{ij...l} \{\q_i\}$ coefficients in the GL functional (see Equation \eqref{eqn:GL} and Methods), therefore allowing the coefficients to be re-expressed purely in terms of geometric quantities. 

Specifically for the quadratic susceptibility, we find that the geometric factor assumes a particularly simple form and is equal to the QGT in the non-occupied subspace $\bar{S}_\kv$ projected onto the flatband subspace at a quasi-momentum $\q$ away (summed over all $\kv$). It is also straightforward to see from Equation \ref{eqn:chi_geo} that this geometric enhancement is maximized when $P_\kv \bar{\mathcal{Q}}_{\kv+\q} = \bar{\mathcal{Q}}_{\kv+\q} \ \forall \, \kv$; in other words, we find that the system is most unstable to ordering if the Bloch bundle at $\kv$ rotates into the Bloch bundle at $\kv+\q$. Formally, we require that the subspace $V$ in $\bar{S}_\kv$ that the projector $P_\kv$ rotates into (that is, $V$ spanned by the eigenvectors of $\bar{\mathcal{Q}}_\kv$) is strictly a subset of $S_{\kv+\q}$. A more careful analysis involving sub-leading order terms (see Methods) additionally requires $\partial_\kv P_\kv = -\partial_{\kv+\q} P_{\kv+\q}$ for $\chi(\q)$ to truly saturate its upper bound. Importantly, we note that these, being local criteria in $\kv$, are not well-described by the global (integrated over $\kv$) geometric nesting criteria defined in Refs. \cite{kivelson, balents}. Particularly, in real materials where the Hamiltonians and hence, the quantum geometry can deviate significantly from ideal scenarios, the conditions defined here, if satisfied for an extended subspace of the Brillouin Zone, will drive a CDW transition even if they are not globally satisfied (akin to Fermi surface nesting). Additionally, since they emerge naturally as modifications to the Ginzburg-Landau free energy coefficients, quantitative conclusions can be made both for systems where the aforementioned upper bound is not saturated and where inclusion of higher order terms (beyond the quadratic susceptibility in Ref. \cite{balents}) becomes necessary, thereby allowing for a more complete theory replete with an exact description of the order parameter as well. 

Local criteria in $\kv$ (as derived here) become even more important in quasi-flatband systems (such as the TMDs considered here) where $k_B T \ll W \ll E_g$, that is, when the system has a well-defined Fermi surface. Here, only states at or near the Fermi surface are relevant (quantitatively chosen by the Lindhard susceptibility), and ordering criteria necessarily need to be local in $\kv$. We find that strong variations of orbital character at the Fermi energy can result in large $\bm v_2(\kv)$. Particularly, it can be shown (see Methods) that to leading order in such systems, $\bm v_2(\kv) \sim - E_g \partial_\kv P_\kv$, and thus the strength of $\partial_\kv P_\kv$ at the Fermi energy (where the bare Lindhard susceptibility is largest) determines the system's instability to charge ordering. Explicitly, the same conditions as in the perfectly flat scenario must be satisfied here, but only within a region $\sim \pm k_B T$ of the chemical potential (see Figure \ref{fig:fig1}). Intuitively, these translate to suggesting that geometric enhancements are most pronounced when the occupied states at $\kv$ acquire the flavor character of unoccupied states at $\kv+\q$ near the Fermi energy. Additionally, this effect is strongest when the states at $\kv$ and $\kv+\q$ completely exchange flavor character near the Fermi energy (for example, at band inversion points). Note that this condition is distinct from the two points already having similar flavor character since $P_\kv \sim P_{\kv+\q} \Rightarrow \chi(\q) \rightarrow 0$. 

We also derive universal commensurability criteria grounded in the microscopics of a material. Specifically, we note that at commensurate $\Q$-vectors, the free energy can receive additional contributions from terms where the $\Q$-vectors add to a reciprocal lattice vector. Depending on whether such additional contributions are in-phase or out-of-phase with the incommensurate contributions, a commensurability-incommensurability transition can be predicted (see Figure \ref{fig:fig3}). These extra terms also lock the phase of the order parameter, leading to the gapping of the phason Goldstone bosons. 

While this work is focused on long-standing questions on CDW formation in imperfectly nested systems, the ideas behind the analysis here are easily generalizable to other phenomena too. For example, we are working on decoupling the interaction term in Equation \ref{eqn:H_int} in the pairing channel to determine geometric contributions to phonon-mediated superconductivity and pair-density wave formation. Since the TMDs considered here also exhibit superconductivity at low temperatures, an analysis of the competition between the pairing and density channel decouplings can reveal novel insights into the interplay between CDW formation and superconductivity hypothesized from experiment~\cite{Morosan2006cdwsc,freitas2016cdwsc,Neupert2022cdwsc}. Additionally, we expect the superconducting gap to inherit the strong $\kv$-dependence of the $\bm \lambda^{mn}_{\kv\kv'}$, thus providing insights into topological superconductivity as well. It would also be interesting to investigate how the addition of a Coulomb potential changes the form of the $\bm \lambda^{mn}_{\kv\kv'}$ factors and whether the system can be driven into a strongly correlated phase such as a Wigner crystal. Magnetic interaction terms also pose intriguing questions; here, the spin-spin order parameter (being dependent on Cartesian indices) acquires even more structure, and exotic magnetic textures may be investigated with associated geometric enhancements. Additionally, if these order parameters transform in different irreducible representations, the resulting Ginzburg-Landau theory could have important topological effects too. 

Our work is qualified by the applicability of the electron-phonon matrix element assumed in ref. \cite{varma}. Generally, $\bm \lambda_{\kv\kv'}^{mn} = \int \psi_{m\kv}^* \nabla \mathcal{U} \psi_{n\kv'} \, d\rv$ is the electronic matrix element of the gradient of the crystal potential $\mathcal{U}(\rv)$ \cite{mcmillan1968lambda}. It is related to the standard electron-phonon coupling constant $\lambda$ in Eliashberg theory \cite{Eliashberg1960} by an average over the Fermi surface $\lambda \propto \langle \bm \lambda ^2 \rangle$ \cite{mcmillan1968lambda}. It can be argued that its equivalence with the difference of velocity operators is at leading order and is a good approximation for orbitals that are well-localized on the atomic sites (such as the $d$ orbitals in transition metal compounds) \cite{varma}. Alternatively, it can be derived exactly by employing a Gaussian approximation for the coupling strength as a function of distance, an assumption that is well-justified for $s$ orbitals as well as orbitals with $s$-character in the dimensionality of the material (such as $p_z$ orbitals in 2D) \cite{yu2024epc.geo}. In general, it can be expected to hold wherever angular contributions to the gradient of the crystal potential are either negligible or have the same dependence as the radial contributions to leading order.

\sect{Conclusions}
To summarize, we find that charge ordering in imperfectly nested systems can be driven by higher order terms and geometric enhancements at specific $\Q$-vectors. We apply our theory to TMDs where standard bare Lindhard susceptibility calculations exhibit no strong dependence on ordering wavevectors and are thus found to be insufficient to explain experimentally-found charge order near $2/3 \, \G \text{M}$. We also predict commensurability transitions in such materials derived from the microscopic Hamiltonians. More generally, our theory can be applied to all systems where phononic effects play dominant roles and where existing calculations based on band energetics alone are found to be insufficient. We discover that systems where the Bloch orbital character rotates rapidly near the Fermi energy (thus imbuing non-trivial quantum geometry) are more susceptible to ordering. Finally, we note that analyses in a similar vein can be used to derive corresponding geometric conditions for other phenomena such as superconductivity and magnetic ordering as well.

%% METHODS %%%

\section*{Methods}

\subsection*{Ginzburg-Landau functional}

We construct the imaginary-time partition function for electronic fields $\psi_{m,k}$ in the presence of phonon-mediated interactions (as described by $H_{\text{int}}$ in Equation \eqref{eqn:H_int}) 

\begin{equation}
    \mathcal{Z} = \int \mathcal{D}[\bar{\psi},\psi] \exp \left( -S_\text{el} + \frac{T^3}{4} \sum_q \bm\rho_q g_q \bm\rho_{-q} \right),
\end{equation}

\noindent where $S_{\text{el}} = T\sum_{nk}\bar{\psi}_{n,k}(-i\omega_p+\xi_{n,\kv})\psi_{n,k}$, and the geometric $\bm \lambda^{mn}_{\kv\kv'}$ factors are hidden in $\bm\rho_q = \sum_k \bm\lambda^{mn}_{\kv+\q,\kv} \bar{\psi}_{m,k+q} \psi_{n,k}$. The summations are carried out over $k=(\omega_p,\kv)$ and $q=(\omega'_p,\q)$, where $\omega_p=\pi(2p+1)T$ are the fermionic Matsubara frequencies and $\omega'_p=2\pi pT$ are the bosonic Matsubara frequencies.

Introducing a Hubbard-Stratonovich decoupling field $\bm\phi_q$ in the density-density channel, the partition function can be rewritten as 

\begin{equation}
    \mathcal{Z} = \int \mathcal{D} [\bar{\psi},\psi,\bm\phi] \exp \left( -S_\text{el} - S_{\text{int}}\right),
\end{equation}

\noindent where $S_{\text{int}} = 1/T \sum_q \bar{\bm\phi}_q g_q^{-1} \bm\phi_q - T \sum_q \bm\rho_{-q} \cdot \bm\phi_q$. Note that $\bm\phi_q$ inherits the structure of $\bm\rho_q$ (to which it is coupled) such that $\bar{\bm\phi}_q=\bm\phi_{-q}$.

Integrating out the fermionic degrees of freedom now, the effective action can be rewritten as

\begin{equation} \label{eqn:S_eff}
    S_{\text{eff}}[\bm\phi] =  \frac{1}{T} \sum_q \bar{\bm\phi}_q g_q^{-1} \bm\phi_q - \text{Tr} \ln \mathcal{G}^{-1},
\end{equation}

\noindent where the electronic Green's function $\mathcal{G}$ can be expressed using the non-interacting Green's function $\mathcal{G}_{0,(mk),(nk')}=[T(-i\omega_p+\xi_{n,\kv})]^{-1}\delta_{mn}\delta_{kk'}$ as

\begin{align}
    \mathcal{G}^{-1}_{(mk),(nk')} &:= (\mathcal{G}_0^{-1} + \hat{\Phi})_{(mk),(nk')} \\ &~= \mathcal{G}_{0,(mk),(nk')}^{-1} + T\sum_q \bm\lambda^{mn}_{\kv,\kv'} \cdot \bm\phi_{k'-k}.
\end{align}

\begin{figure*}
    \centering
    \includegraphics[width=0.95\linewidth]{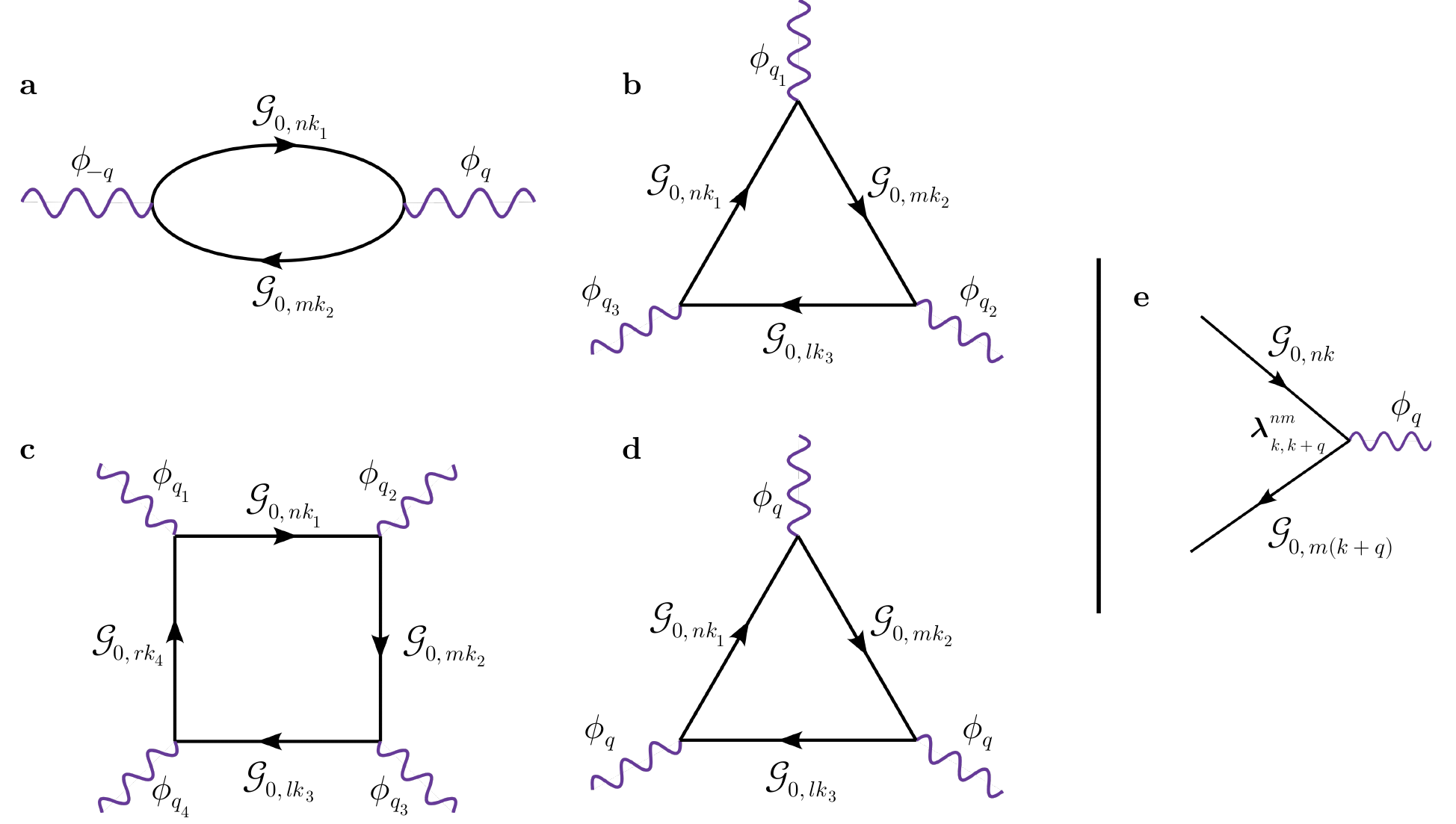}
    \caption{\textbf{Diagrammatic representations of GL coefficients.} The effective action and hence the GL free energy functional affords diagrammatic representations as shown of the electron-phonon interaction at \textbf{a} quadratic, \textbf{b} cubic, and \textbf{c} quartic order. \textbf{d} At commensurate wavevectors (for example, $2/3 \, \G \text{M}$ in the TMDs considered here), the cubic term involves an extra contribution from umklapp scattering, whose phase interference with the normal cubic term in \textbf{(b)} determines a commensurability transition. \textbf{e} All terms involve interaction vertices of this form, where a bare electronic state in band $n$ and wavevector $\kv$ is scattered by a phonon of wavevector $\q$ into a state in band $m$ and wavevector $\kv+\q$. The dependence on electronic geometry is encoded in the interaction vertex $\bm \lambda^{nm}_{\kv,\kv+\q}$. Note that the diagonal nature of the bare Green's function has been used to concisely represent $\mathcal{G}_{0,(mk),(nk')}$ as $\mathcal{G}_{0,mk}$.}
    \label{fig:loops}
\end{figure*}

Near the CDW transition, we expect $\bm\phi_q$ to be small and can expand the action in powers of $\bm\phi_q$ to obtain the Ginzburg-Landau functional. Writing the second term in Equation \eqref{eqn:S_eff} as $-\text{Tr} \ln (\mathcal{G}_0^{-1}(\mathcal{I}+\mathcal{G}_0\hat{\Phi}))=-\text{Tr} \ln \mathcal{G}_0^{-1} + \sum_j(-1)^j\, \text{Tr}(\mathcal{G}_0\hat{\Phi})^j/j$, and ignoring the term independent of $\phi$, the effective action can be written as

\begin{equation}
    S_{\text{eff}}[\bm\phi] = \frac{1}{T} \sum_q \frac{1}{g(\q)} |\phi_q|^2 + \sum_j\frac{(-1)^j}{j}\, \text{Tr}(\mathcal{G}_0\hat{\Phi})^j.
\end{equation}

Diagrammatically, $\hat{\Phi}$ connects bare electronic propagators $\mathcal{G}_0$ at different wavevectors, and the associated diagrams for the quadratic, cubic and quartic terms are shown in Figure \ref{fig:loops}. 

Within a mean-field description and in the static limit (setting $\omega'_p=0$), Equation \eqref{eqn:GL} for the free energy is obtained with the coefficients $f^{(n)}\{\q_i\}$ given by

\begin{equation} \label{eqn:fn}
    f^{(n)}\{\q_i\} = \frac{1}{n} \sum_{\{m_i\},\, \kv} L^{(n)}\{\xi_{m_i,\kv_i}\} \bigotimes_{i}^n\bm\lambda^{m_im_{i+1}}_{\kv_i,\kv_{i+1}},
\end{equation}

\noindent where the $\kv_i=\kv+\sum_j^{i-1}\q_j$ form an $n$-sided polygon in the BZ (as earlier, $\sum_i^n \q_i = \vec{G}$), and $m_{n+1}=m_1$. A Matsubara summation has been carried out over all fermionic frequencies $\omega_p$ to obtain generalized bare susceptibilities $L^{(n)}\{\xi_{m_i,\kv_i}\} = \sum_i^n n_F(\xi_{m_i,\kv_i})/\prod_{j \neq i} (\xi_{m_i,\kv_i}-\xi_{m_j,\kv_j})$. Note that for $n=2$, $-f^{(2)}$ reduces to the $\chi(\q)$ in Equation \eqref{eqn:chi}, and $L^{(2)}$ reproduces the standard Lindhard susceptibility. While Equation \eqref{eqn:fn} may in general give complex $f^{(n)}$ for a particular set $\{\q_i\}$, the GL expansion contains a term formed from the set $\{-\q_i\}$ too, which is the complex conjugate of the former and thus always results in real coefficients as expected.

\subsection*{Flatband limit}

While Equation \eqref{eqn:fn} is general, we elaborate on the specific flatband limit where it considerably simplifies. Particularly, we investigate the scenario when the chemical potential lies within a set of $N_{\text{flat}}$ bands (called flatbands) which are well-separated in energy from all other bands. Explicitly, their bandwidth $W$ is much smaller than the band gap $E_g$. For simplicity, we also assume that the thermal scale in the system is such that these $N_{\text{flat}}$ bands are roughly equally occupied at all $\kv$ (this assumption shall be relaxed later), and excitations across the band gap are negligible. Effectively, we work within the limit such that $W \ll k_BT \ll E_g$. As a result, the generalized bare susceptibilities  $L^{(n)}\{\xi_{m_i,\kv_i}\}$ become approximately constant over $\kv$, and at leading order, only contain contributions from bands $m_i \in [1, N_\text{flat}]$ (owing to the energy differences in the denominator, any other band contributions is suppressed by a factor of $W/E_g$). The Ginzburg-Landau coefficients thus simplify to 

\begin{equation}
    f^{(n)}\{\q_i\} \approxprop  \sum_\kv \text{Tr} \, \widehat{\bigotimes}_i^n(P_{\kv_i}\bm\Lambda_{\kv_i,\kv_{i+1}}),
\end{equation}

\noindent where $\bm\Lambda_{\kv,\kv',\alpha\beta}=U_{\alpha m}(\kv)\bm\lambda^{mn}_{\kv,\kv'}U_{\beta n}^{\dagger}(\kv')$ and $\widehat{\otimes}$ denotes normal matrix multiplication over flavor indices but a tensor product over Cartesian indices. The trace is over flavor indices as well, and as before, $\kv_i=\kv+\sum_j^{i-1}\q_j$. 

In the small bandwidth limit, contributions to $\bm\lambda^{mn}_{\kv,\kv'}$ are dominated entirely by the geometric part of the velocity operator $\bm v_2 (\kv)=\sum_n E_n(\kv)\partial_\kv P^n(\kv)$. If we additionally assume that bands which share flavor character with the flat bands have bandwidth much smaller than $E_g$ (the generalization to multiple well-separated subspaces of bands is just additive in the respective energy gaps), $\partial_\kv H_0(\kv) \approx E_g \partial_\kv \bar{P}_\kv$, and $\bm\lambda^{mn}_{\kv,\kv'}$ reduces to the expression given earlier involving differences of projector derivatives. Importantly, this allows the GL coefficients to be rewritten purely in terms of geometric quantities as 

\begin{equation}\label{eqn:fn_geo}
    f^{(n)}\{\q_i\} = (i E_g)^n \,\sum_\kv\text{Tr} \, \widehat{\bigotimes}_i^n\left[P_{\kv_i} (\nabla_{\kv_i} P_{\kv_i} - \nabla_{\kv_{i+1}} P_{\kv_{i+1}}) P_{\kv_{i+1}}\right].
\end{equation}

For $n=2$, Equation \eqref{eqn:fn_geo} results in two types of geometric terms, one of which is given in Equation \eqref{eqn:chi_geo}. The other term is proportional to $-1/2\sum_\kv (P_\kv \nabla_\kv \bar{P}_\kv + P_{\kv+\q}\nabla_{\kv+\q}\bar{P}_{\kv+\q})^{\widehat{\otimes}2} + \text{h.c.}$, which in general requires both terms inside the brackets to be significant (i.e., it requires the off-diagonal $S_\kv \rightarrow \bar{S}_\kv$ block of the non-Abelian Berry connection to be non-negligible at both $\kv$ and $\kv+\q$). Based on a degree-of-freedom counting analysis then, we argue that the QGT term in Equation \eqref{eqn:chi_geo} is the dominant contribution to $\chi(\q)$. Note that an additive contribution from these sub-leading order terms can be ensured if $\text{sgn}\left(\nabla_\kv \bar{P}_\kv\right)=-\text{sgn}\left(\nabla_{\kv+\q}\bar{P}_{\kv+\q}\right)$, which reproduces the conditions noted earlier for $\chi(\q)$ to saturate its upper bound. The higher order terms can similarly be expanded in terms of projectors and their derivatives at different $\kv$-points; they do not, however, afford a closed form expression in terms of the QGT as in Equation \eqref{eqn:chi_geo}.

In quasi-flatband systems, $W \ll k_BT$ is no longer a good approximation. However, the results from this section, albeit with some modifications, still apply. Working instead in the $k_BT \ll W \ll E_g$ limit, the generalized bare susceptibilities $L^{(n)}\{\xi_{m_i,\kv_i}\}$ are no longer constant over $\kv$, but are instead sharply peaked at the Fermi surface (with width in $\kv$-space given by $\sim k_BT/Wa \ll 1/a$). The summation over $\kv$ in Equation \eqref{eqn:fn} then effectively reduces to a sum over the Fermi surface multiplied by some constant. Equation \eqref{eqn:fn_geo} is similarly modified with the summation now running over $\kv$ along the Fermi surface, along with an overall multiplicative scaling arising from the strength of $L^{(n)}$ at the relevant $\q$-vectors. Importantly, if $L^{(n)}$ is relatively flat as a function of $\q$, CDW formation is driven entirely by the same geometric factors derived in this section - this is the case for the TMDs considered in this work, where Figure \ref{fig:fig2} highlight the importance of the geometric factors in selecting the CDW $\Q$-vectors.

\subsection*{Hexagonal lattices}

In this section, we derive the coefficients $f^{(n)}$ for hexagonal lattices where we get triple-$\Q$ charge order and show how the tensorial GL expansion simplifies to the scalar version in Equation \eqref{eqn:F_tripleQ}. Explicitly looking for such solutions effectively requires expressing $\mathcal{F}_{\Q_j}$ as dependent only on $\bm\phi_{\Q_j}$ (for $\Q_j=R_{2\pi j/3} \Q_0$) and minimizing it to obtain the ordering vectors $\Q_p$. The allowed terms $\mathcal{F}^{(n)}$ at any order $n$ in the GL expansion then become heavily constrained by the condition $\sum_i^n \q_i = \vec{G}$, and the free energy up to quartic order for incommensurate $\Q_0$ can be written as 

\begin{equation}
    \mathcal{F}_{\Q_0} = \sum_j \mathcal{F}^{(2)}_{\pm\Q_j} + \left(\mathcal{F}^{(3)}_{\Q_0, \Q_1, \Q_2} + \text{c.c.}\right) + \sum_{i,j}\mathcal{F}^{(4)}_{\pm\Q_i,\pm\Q_j},
\end{equation}

\noindent where the dependence of each $\mathcal{F}^{(n)}$ on the $n$ $\Q$-dependent order parameters is made explicit. For commensurate $\Q_0$ such that $3\Q_0 = \vec{G}$, there are additional cubic $\sum_j \mathcal{F}^{(3)}_{\Q_j,\Q_j,\Q_j}+ \text{c.c.}$ and quartic $\sum_j\mathcal{F}^{(4)}_{\Q_j, \Q_j, -\Q_{j+1},-\Q_{j+2}}+ \text{c.c.}$ terms.

The $C_3$ symmetry of the lattice forces $\mathcal{F}_{\Q_0}$ to transform under the trivial irrep (irreducible representation) and thus imposes additional structure on the order parameter. Specifically, owing to the $C_3$ symmetry of the $H_0 (\kv)$, each coefficient $f^{(n)}$ transforms in the trivial irrep ($f^{(n)} \{\q_i\} \rightarrow (R_{2\pi j/3}^{-1})^{\otimes n} f^{(n)} \{R_{2\pi j/3}\q_i\}=f^{(n)} \{\q_i\}$); consequently, the objects depending on the order parameter in $\mathcal{F}^{(n)}$ must also transform in the trivial irrep. For example, for the quadratic term, this object is the covariant tensor $\bm\phi_{\Q_j}\otimes\bar{\bm\phi}_{\Q_j}$, and transformation in the trivial irrep implies $\bm\phi_{R_{2\pi j/3}\Q_j}\otimes\bar{\bm\phi}_{R_{2\pi j/3}\Q_j}=(R_{2\pi j/3}\bm\phi_{\Q_j})\otimes (R_{2\pi j/3}\bar{\bm\phi}_{\Q_j})$.  Importantly, this leaves the overall phase of the order parameter across different $\Q_j$ unconstrained and allows it to be written as $\bm\phi_{\Q_j} = e^{i\alpha_{\Q_j}}\Delta_{\Q_0} R_{2\pi j/3} \hat{\bm n}_{\Q_0}$ (the corresponding cubic and quartic objects do not reduce the degrees of freedom of $\bm\phi$ any further). For commensurate $\Q_0$, $\bm\phi_{\Q_j}^{\otimes3}$, and hence, $\bm\phi_{\Q_j}$ also transforms in the trivial irrep under $C_3$. The phases thus become constrained ($\alpha_{\Q_j}=\alpha_{\Q_0} \  \forall j$), and two degrees of freedom in the order parameter are lost corresponding to the gapping of the two phason modes during the commensurability transition. 

Premultiplying the tensorial coefficients $f^{(n)}\{\Q_j\}$ with tensor products of the relevant rotation matrices $R_{2\pi j/3}^{-1}$, the free energy can be expressed in terms of contractions with just the vectors $\hat{n}_{\Q_0}$. Suppressing the $\Q_0$ dependence henceforth for clarity, the free energy up to quartic order is 

\begin{equation}
    \mathcal{F}=a \hat{\bm n}^{\otimes2} \Delta^2 + \text{Re}\left(c\hat{\bm n}^{\otimes 3} \Delta^3 e^{i\sum_{j=0}^2\alpha_j}\right) + b\hat{\bm n}^{\otimes 4} \Delta^4,
\end{equation}

\noindent where $a$, $c$ and $b$ are totally symmetric rank-$2$, $3$, and $4$ tensors dependent on the $f^{(n)}$ coefficients and are contracted with $\hat{\bm n}$ above. Identifying $\alpha_{j}=\alpha_{0} \  \forall j$ reproduces the form for the commensurate case. Diagonalizing $a$ to get eigenvalues $\lambda_i$ and eigenvectors $\bm e_{i,0}$, the tensors $c$ and $d$ can be re-expressed in the eigenbasis of $a$. The order parameter then in this basis becomes $\bm\phi_{j} = \left(\varphi \bm e_{1,0} + \kappa\bm e_{2,0} \right)e^{i\alpha_j}$. 

Since the system is likely to exhibit a preference for condensing along a particular phonon polarization, the free energy is expected to be highly anisotropic in the $\varphi$-$\kappa$ space (see SI for evidence regarding the same in the TMDs considered here). Concretely, we assume that near the transition temperature, one of the eigenvalues of $a$ is much smaller than the other (without loss of generality, $\lambda_1 \ll \lambda_2$). Along the $\Gamma \text{M}$ line in the TMDs studied here, $f^{(2)}\{\pm\q\}$ was explicitly found to be highly anisotropic with a preference for the longitudinal polarization. Then, it is expected that the order parameter $(\varphi_\star, \kappa_\star)$ that minimizes $\mathcal{F}$ must be along the $\varphi$-axis to leading order, that is, $\kappa_\star$ is at least $O(\varphi_\star^2)$ (this claim will be self-consistently verified later). Up to quartic order in $\varphi$, the free energy reduces to 

\begin{equation}\label{eqn:F_phikappa}
    \mathcal{F}=\lambda_1 \varphi^2 - \tilde{c}_{111} \varphi^3 + \left(b_{1111}\varphi^4 - 3\tilde{c}_{112}\varphi^2\kappa + \lambda_2 \kappa^2\right),
\end{equation}

\noindent where the phase dependence has been absorbed into the cubic coefficients $\tilde{c}=-\text{Re} \, c e^{i\sum_{j=0}^2\alpha_j}$. Minimizing with respect to $\kappa$ reveals $\kappa_\star = 3\tilde{c}_{112}\varphi^2/2\lambda_2$. Substituting back into the free energy, the quartic coefficient becomes $b'_{1111}=b_{1111}-9\tilde{c}_{112}^2/4\lambda_2$, and the global minimum shifts from $\varphi_\star=0$ (disordered) to $\varphi_\star=\tilde{c}_{111}/2b_{1111}'$ (ordered) when $\lambda_1=\tilde{c}_{111}^2/4b_{1111}'$. The off-axis component of the minima is $\kappa_\star = 3(\lambda_1/\lambda_2 )\tilde{c}_{112}\varphi_\star/\tilde{c}_{111} \ll \varphi_\star$ since $\lambda_1/\lambda_2 \ll 1$ near the transition, thus verifying the earlier claim. 

Writing $\Delta=\sqrt{\varphi^2+\kappa^2}\approx \varphi \left(1+9\tilde{c}_{112}^2\varphi^2/8\lambda_2^2\right)$, and comparing the free energy coefficients in Equation \eqref{eqn:F_phikappa} with Equation \eqref{eqn:F_tripleQ}, one can identify $\left(3/\text{Tr} \, g - \Pi\right) = \lambda_1$, $\gamma=c_{111}$, and $\beta=b_{1111}'$ (where for the quartic term, we use $\lambda_1 \ll \lambda_2)$. Importantly, the transition condition in Equation \eqref{eqn:transition} is recovered from $\lambda_1=\tilde{c}_{111}^2/4b_{1111}'$, and the explicit dependence of the coefficients on the microscopic $f^{(n)}$ terms can be written for incommensurate $\Q_0$ as

\begin{align}
    \Pi_{\Q_0} &= 6\, \lambda_{\text{min}} \left[ f^{(2)} \{\pm \Q_0\} \right], \\
    \gamma_{\Q_0} &= 2 \sum_{a\neq b\neq c} f^{(3)}_{ijk}\{\Q_a,\Q_b,\Q_c\}\bm e_{1,a}^i\bm e_{1,b}^j\bm e_{1,c}^k, \\
    \beta_{\Q_0} &= \sum_{\substack{\text{perm.} \\ a \leq b}} f^{(4)}_{ijkl} \{\pm\Q_a,\pm\Q_b\} \bm e_{1,a}^i\bm e_{1,a}^j\bm e_{1,b}^k\bm e_{1,b}^l, 
\end{align}

\noindent where $\bm e_{1,0}$ is the eigenvector corresponding to the minimal eigenvalue of $f^{(2)} \{\pm \Q_0\}$ in the expression for $\Pi_{\Q_0}$, and $\bm e_{1,a} = R_{2\pi a/3}\bm e_{1,0}$. The sums run over permutations of $\{\q_i\}$ in $f^{(n)}$. For commensurate $\Q_0$, $\gamma_{\Q_0} \rightarrow \gamma_{\Q_0}+\gamma'_{\Q_0}$ and $\beta{\Q_0} \rightarrow \beta_{\Q_0}+\beta'_{\Q_0}$, where the additional terms are given by

\begin{align}
    \gamma'_{\Q_0} &= 6 \, f^{(3)}_{ijk}\{\Q_0,\Q_0,\Q_0\}\bm e_{1,0}^i\bm e_{1,0}^j\bm e_{1,0}^k, \\
    \beta'_{\Q_0} &= \text{Re}\sum_{\substack{\text{perm.} \\ a\neq b \neq c}} f^{(4)}_{ijkl} \{\Q_a,\Q_a, -\Q_{b},-\Q_{c}\} \bm e_{1,a}^i\bm e_{1,a}^j\bm e_{1,b}^k\bm e_{1,c}^l.
\end{align}

\noindent In the above, it has been additionally assumed that the interaction vertex $g_\q$ is itself strongly anisotropic, such that $\bm e_{2,0}^T g^{-1}_{\Q_0} \bm e_{2,0} \gg \tilde{c}_{112}^2/b_{1111}$ so that the hard mode (from $\kappa_\star$) corrections to the quartic coefficient derived earlier are negligible. If this is not the case, $\beta_{\Q_0}$ also contains contributions from $\lambda_2$ and contractions of $f^{(3)}$ with $\bm e_{2,0}$. 

\subsection*{First-principles Calculations}

First-principles electronic-structure calculations for monolayer NbSe$_2$, TaSe$_2$, and TaS$_2$ were performed within Density Functional Theory using the Perdew--Burke--Ernzerhof (PBE) exchange-correlation functional~\cite{perdew1996generalized} and the projector augmented-wave (PAW) formalism~\cite{blochl1994projector} as implemented in \texttt{VASP}~\cite{kresse1996efficient,kresse1999ultrasoft}. Relativistic effects, including spin-orbit coupling, were fully taken into account. For all three materials, the plane-wave kinetic-energy cutoff was set to 400~eV and the electronic self-consistency threshold was set to $10^{-7}$~eV. The Brillouin zone was sampled using a $12\times12\times1$ $k$-mesh.

%For NbSe$_2$, periodic images of the monolayer were separated using an out-of-plane lattice parameter of 25.094~\AA. For TaSe$_2$, periodic images were separated using an out-of-plane lattice parameter of 32.130~\AA. For TaS$_2$, periodic images were separated using an out-of-plane lattice parameter of 30.000~\AA.

To calculate the generalized susceptibility coefficients, a six-band (10-band) tight-binding Hamiltonain was constructed for NbSe$_2$ (TaS$_2$ and TaSe$_2$) using \texttt{Wannier90}~\cite{mostofi2008wannier90} with Nb $4d_{xy}$, $4d_{x^2-y^2}$, and $4d_{z^2}$ (all Ta-$5d$) orbitals as the projection centres. 
%These Wannier Hamiltonians were used to calculate the generalized susceptibility coefficients entering the charge-density-wave driver.

% %% DECLARATIONS %%%
% \bigskip

% %% DATA AVAILABILITY %%%
% \section*{Data Availability}
% The datasets generated and analyzed in this study are available from the corresponding author upon request.

% %%% CODE AVAILABILITY %%%
% \section*{Code Availability}
% The data presented in this study were generated using theoretical models as well as free and open-source first-principles packages as described in the Methods section. 

%%% ACKNOWLEDGMENTS %%%
\section*{Acknowledgments}
A.~A.~acknowledges funding from the Cambridge International Scholarship awarded by the Cambridge Trust.
R.-J.S. acknowledges funding from an EPSRC ERC underwrite Grant No.~EP/X025829/1. R.G. and M.S.B. gratefully acknowledge the Center for Computational Materials Science at Institute for Materials Research (IMR) for allocations on the MASAMUNE-IMR supercomputer system (Project No. 202112-SCKXX-0510). M.S.B. acknowledges support from Leverhulme Trust (Grant No. RPG-2023-253).

%%% AUTHOR CONTRIBUTIONS %%%
\section*{Author Contributions}
R.-J.~S. and A.~A. conceptualized the idea and theory, where A.~A. developed the framework under the supervision of \mbox{R.-J.~S.} A.~A. and R.-J.~S. were informed on material candidates by M.~S.~B. R.~G. carried out the first-principles calculations under the guidance of  M.~S.~B., and A.~A. performed the Ginzburg-Landau calculations on all these models with inputs from R.~G. All authors contributed to the analysis and interpretation of the results. A.~A. and R.-J.~S. wrote the manuscript with input from all authors.

\clearpage
\onecolumngrid
\setcounter{equation}{0}\setcounter{figure}{0}
\setcounter{table}{0}\setcounter{section}{0}
\renewcommand{\theequation}{S\arabic{equation}}
\renewcommand{\thefigure}{S\arabic{figure}}
\renewcommand{\thetable}{S\arabic{table}}
\renewcommand{\thesection}{S\arabic{section}}

\begin{center}
  {\large\bfseries SUPPLEMENTARY INFORMATION \\[0.3em]
  Geometric Ginzburg-Landau theory of charge ordering and commensurability\\[1em]}
  Aneesh Agarwal,$^1$ Rutvij Gholap,$^2$ Mohammad Saeed Bahramy,$^2$ and Robert-Jan Slager$^{2,1}$\\[0.5em]
  {\itshape\small $^{1}$Theory of Condensed Matter Group, Cavendish Laboratory, University of Cambridge, J. J. Thomson Avenue, Cambridge CB3 0HE, United Kingdom\\
  $^2$Department of Physics and Astronomy, University of Manchester, Oxford Road, Manchester M13 9PL, United Kingdom}
\end{center}
\vspace{1.5em}

\section*{Supplementary Note 1: Summary of Ginzburg-Landau coefficient calculations}

We present our calculations of the individual Ginzburg-Landau (GL) coefficients highlighting how the triangular scattering process in the cubic term is essential in selecting the ordering wavevector. The integrals were calculated on a $1080 \times 1080 \ \kv$-grid with an adaptive mesh refining the coarse grid by a factor of 8 near the Brillouin Zone (BZ) peaks (convergence was tested using a refinement factor of 4, and the coefficients agreed to within a couple of percent). 

\begin{figure}[p]
    \centering
    \includegraphics[width=\linewidth]{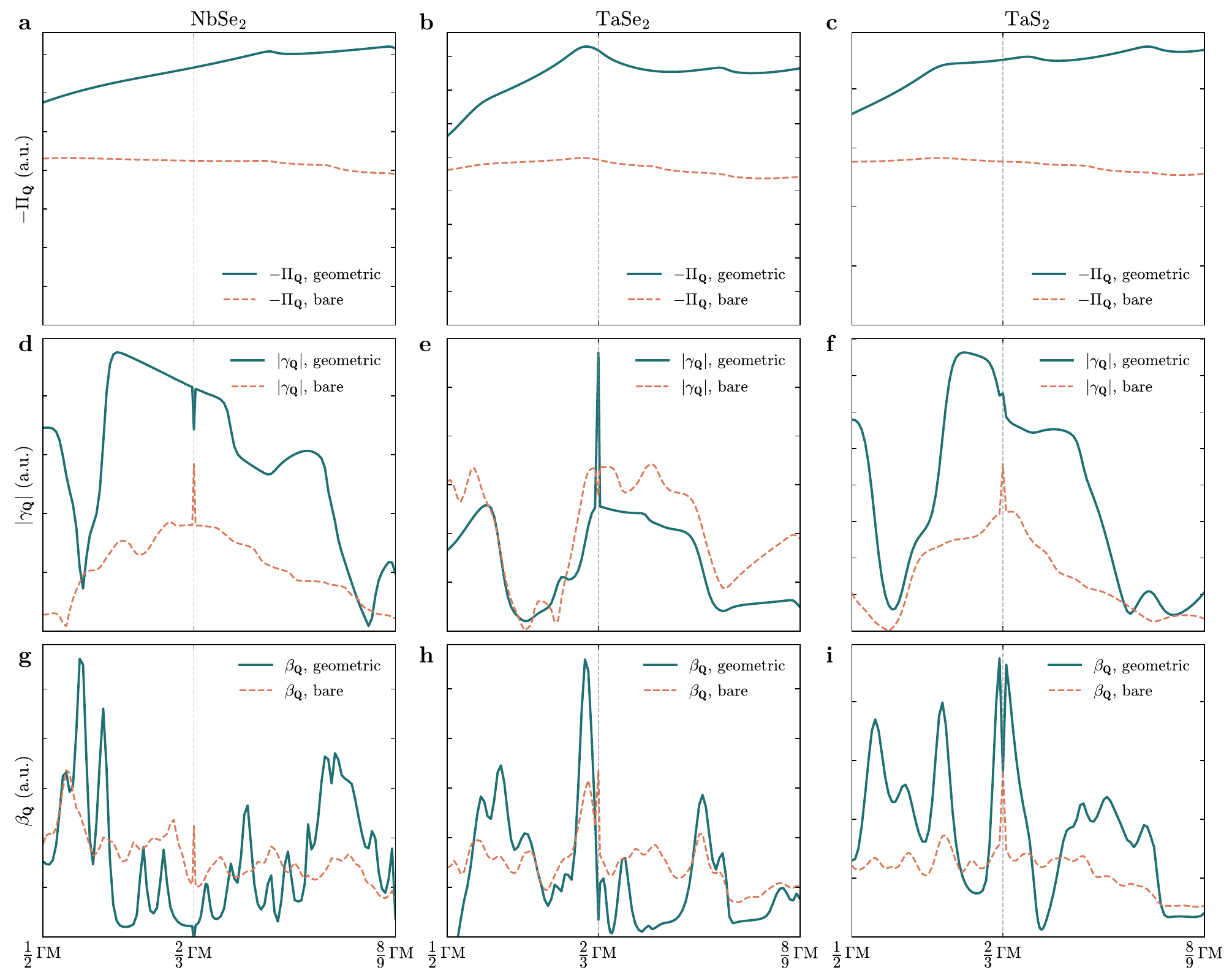}
    \caption{\textbf{a-c} Quadratic $\Pi_\Q$, \textbf{d-f} cubic $|\gamma_\Q|$ and \textbf{g-i} quartic $\beta_\Q$ coefficients as a function of $\Q$ for NbSe$_2$ ($T\approx 23 \, \text{K}$), TaSe$_2$ ($T\approx 35 \, \text{K}$), and TaS$_2$ ($T\approx 49 \, \text{K}$). The bare quadratic susceptibility curve exhibits the broad peak that is commonly seen in literature \cite{SM:zhu2015_cdwclassification,SM:johannes_nbse2_2006,SM:Johannes_nesting_2008,SM:flicker_nbse2_2016}, and the geometric enhancements only marginally alter it and cannot explain the wavevector selection seen in experiment. The cubic coefficient, however, shows a sharp increase near $2/3 \, \G \text{M}$ for all three materials corresponding to ordering wavevectors connecting three points on the Fermi surface. The bimodal nature of the peak reflects the aforementioned being satisfied for both bands at different wavevectors. All three materials exhibit some sort of minima in their quartic coefficient near where the cubic is maximised, resulting in the CDW driver peaks reported in the main text. Importantly, $\beta_\Q$ is found to be positive for almost the entire domain considered here, a property that is essential for the GL free energy to be bounded from below.}
    \label{fig:qcq}
\end{figure}

As expected from literature \cite{SM:zhu2015_cdwclassification,SM:johannes_nbse2_2006,SM:Johannes_nesting_2008,SM:flicker_nbse2_2016}, the bare quadratic coefficient in Fig. \ref{fig:qcq} exhibits broad peaks over the plotting domain for all three materials. Geometric enhancements also do not show any sharp selective behavior in the wavevector (as expected from the shape of the Fermi surface). The cubic coefficient (see Fig. \ref{fig:qcq}) however demonstrates sharp behavior near $2/3 \, \G \text{M}$ owing to the triangular scattering processes connecting three points on the $K$ pocket Fermi surface (FS). Since such processes occur for the FSes of both of the (two) low-lying bands, the peak in $|\gamma_\Q|$ is bimodal in nature. The shorter peak near $1/2 \, \G \text{M}$ corresponds to scattering between states on the $K$ pocket FS and those on the $\G$ pocket FS. However, owing to both the quadratic coefficient being smaller here, and suppression by the quartic in the expression for the CDW driver $\Theta_\Q$, this peak does not result in any physical consequences. The quartic coefficient (see Fig. \ref{fig:qcq}) also shows strong variation with $\Q$, and through its minima, further refines the position of the peaks in the driver. Importantly, it is found to be positive throughout almost the entirety of the plotted domain and is therefore sufficient in bounding the GL functional from below without requiring higher order terms (see Supplementary Note 3 for a consideration of the scenario when it does become negative). 

% \begin{figure*}[t]
%     \centering
%     \includegraphics[width=\linewidth]{images_SM/quadratic.pdf}
%     \caption{Quadratic coefficient $\Pi_\Q$ as a function of $\Q$ for different materials. The bare susceptibility curve (orange dashed line) exhibits the broad peak that is commonly seen in literature, and the geometric enhancements only marginally alter it and cannot explain the wavevector selection seen in experiment.}
%     \label{fig:quadratic}
% \end{figure*}

% \begin{figure*}[t]
%     \centering
%     \includegraphics[width=\linewidth]{images_SM/cubic.pdf}
%     \caption{Normalized cubic coefficient $|\gamma_\Q|$ as a function of $\Q$ for different materials. For all three materials, a sharp increase is seen for the cubic coefficient near $2/3 \, \G \text{M}$ corresponding to ordering wavevectors connecting three points on the Fermi surface. The bimodal nature of the peak reflects the aforementioned being satisfied for both bands at different wavevectors.}
%     \label{fig:cubic}
% \end{figure*}

% \begin{figure*}[t]
%     \centering
%     \includegraphics[width=\linewidth]{images_SM/quartic.pdf}
%     \caption{Quartic coefficient $\beta_\Q$ as a function of $\Q$ for different materials. All three materials exhibit some sort of minima in their quartic coefficient near where the cubic is maximised, resulting in the CDW driver peaks reported in the main text. Importantly, $\beta_\Q$ is found to be positive for almost the entire domain considered here, a property that is essential for the GL free energy to be bounded from below.}
%     \label{fig:quartic}
% \end{figure*}

%\newpage

\section*{Supplementary Note 2: Comments on the insufficiency of bare higher-order terms}

It is interesting to note that the bare higher-order terms in the CDW driver $\Theta_\Q$ contribute negligibly even where the corresponding geometrically modified terms are of similar order of magnitude to $\Pi_\Q$. The origin for this lies in the functional form of $L^{(3)}\{\xi_{m_i,\kv_i}\}$

\begin{equation}
    L^{(3)}\{\xi_{m_i,\kv_i}\} = \frac{n_F(\xi_{m_1,\kv_1})}{(\xi_{m_1,\kv_1}-\xi_{m_2,\kv_2})(\xi_{m_1,\kv_1}-\xi_{m_3,\kv_3})} + \frac{n_F(\xi_{m_2,\kv_2})}{(\xi_{m_2,\kv_2}-\xi_{m_1,\kv_1})(\xi_{m_2,\kv_2}-\xi_{m_3,\kv_3})} + \frac{n_F(\xi_{m_3,\kv_3})}{(\xi_{m_3,\kv_3}-\xi_{m_1,\kv_1})(\xi_{m_3,\kv_3}-\xi_{m_2,\kv_2})}.
\end{equation}

\noindent The transformation $\xi_{m_i,\kv_i} \rightarrow -\xi_{m_i,\kv_i} \, \forall \{m_i, \kv_i\}$ for $T=0$ (step function distribution) implies $L^{(3)}\{\xi_{m_i,\kv_i}\} \rightarrow - L^{(3)}\{-\xi_{m_i,\kv_i}\}$. Since the largest contributions to the cubic coefficient arise from triangular scattering processes where all three states lie on the Fermi surface (i.e., where locally for small perturbations $\text{d}\kv_i$ near the FS, $\xi_{m_i,\kv_i} = \bm v_{m_i} \cdot \text{d} \kv_i$), the integral for the bare $\gamma_\Q$ necessarily involves summations of the form $L^{(3)}\{\xi_{m_i,\kv_i}\} + L^{(3)}\{-\xi_{m_i,\kv_i}\}$. Since these same contributions vanish, the bare $\gamma_\Q$ is entirely made of sub-leading order contributions from points farther away from the FS and therefore contributes negligibly to the CDW driver. In other words, the integrand for the bare $\gamma_\Q$ is locally odd for all three $\kv$ points near the FS and therefore vanishes there. 

The same reasoning works for all $L^{(n)}\{\xi_{m_i,\kv_i}\}$, where $n$ is odd. At $T=0$, $n_F(\xi_{m_i,\kv_i})$ is the step function $\theta (-\xi_{m_i,\kv_i})$, and the transformation $\xi_{m_i,\kv_i}\rightarrow -\xi_{m_i,\kv_i}$ sends $L^{(n)}\{\xi_{m_i,\kv_i}\} = \sum_i^n \theta(-\xi_{m_i,\kv_i})/\prod_{j \neq i} (\xi_{m_i,\kv_i}-\xi_{m_j,\kv_j}) \rightarrow \sum_i^n (1-\theta(-\xi_{m_i,\kv_i}))/\prod_{j \neq i} (\xi_{m_i,\kv_i}-\xi_{m_j,\kv_j})=-L^{(n)}\{\xi_{m_i,\kv_i}\}$.

\newpage

\section*{Supplementary Note 3: Non-positive $\beta_{\Q}$}

If $\beta_\Q$ is found to be negative or close to zero (as it is at $2/3 \, \G \text{M}$ for NbSe$_2$ and near $1/2 \, \G \text{M}$ for TaSe$_2$), higher-order terms are needed to bound the GL free energy from below. These then also modify the transition condition (and hence the CDW driver) in a meaningful manner, and their evaluation is useful for a complete description of the transition. However, it is possible to make qualitative conclusions about the behaviour near these points from existing calculations. Particularly, the $C_3$-symmetry of the TMDs, along with the large cubic coefficients they exhibit, suggests that near these problematic points, a sextic term may be large enough to bound the GL functional. Assuming that $\beta_\Q$ is negligible, the GL free energy with the inclusion of a sextic term of the form $u_\Q \Delta_\Q^6$ with $u_\Q>0$ affords a closed form solution for the transition given by 

\begin{equation}
    \frac{3}{\text{Tr}\, g_\Q} := \Theta_\Q'= -\Pi_\Q + \frac{3|\gamma_\Q|}{4} \left(\frac{|\gamma_\Q|}{4u_\Q}\right)^{1/3}.
\end{equation}

\noindent $\Theta'_\Q$ here is the analog of the CDW driver studied in the main text. 

While we do not explicitly calculate $u_\Q$ here, we do obtain an approximate lower bound estimate for it using the cubic coefficient. Based on the roughly triangular shape of the Fermi surfaces, we infer that the strongest contributions to $u_\Q$ must arise from the same triangular scattering processes (now at second order) that contributed to $\gamma_\Q$. In particular, where $|\gamma_\Q|$ involved an integral over $\kv$ of some integrand $\mathcal{I}_\Q (\kv)/3$, one of the strongest contributors to $u_\Q$ is $\int_{\text{BZ}} \mathcal{I}_\Q^2 (\kv)/6 \ \text{d}^2\kv$ (the other significant contributor is a hexagonal scattering process particularly relevant for the $\G$ pocket FS, but this is not considered here). Then, by the Cauchy-Schwarz inequality, $u_\Q \gtrsim 3|\gamma_\Q|^2/2$, and $\Theta'_\Q \lesssim -\Pi_\Q + 0.41 |\gamma_\Q|^{2/3}$. Since we expect the transition condition to increase monotonically from the CDW driver $\Theta_\Q$ in the main text to the modified version $\Theta'_\Q$ considered here as $\beta_\Q$ decreases to zero, the approximate upper bound derived here for $\Theta'_\Q$ also applies to $\Theta_\Q$. Further, it can actually be used to signal where an inclusion of sextic terms may be useful even for positive (but small) $\beta_\Q$. 

\begin{figure}[h]
    \centering
    \includegraphics[width=\linewidth]{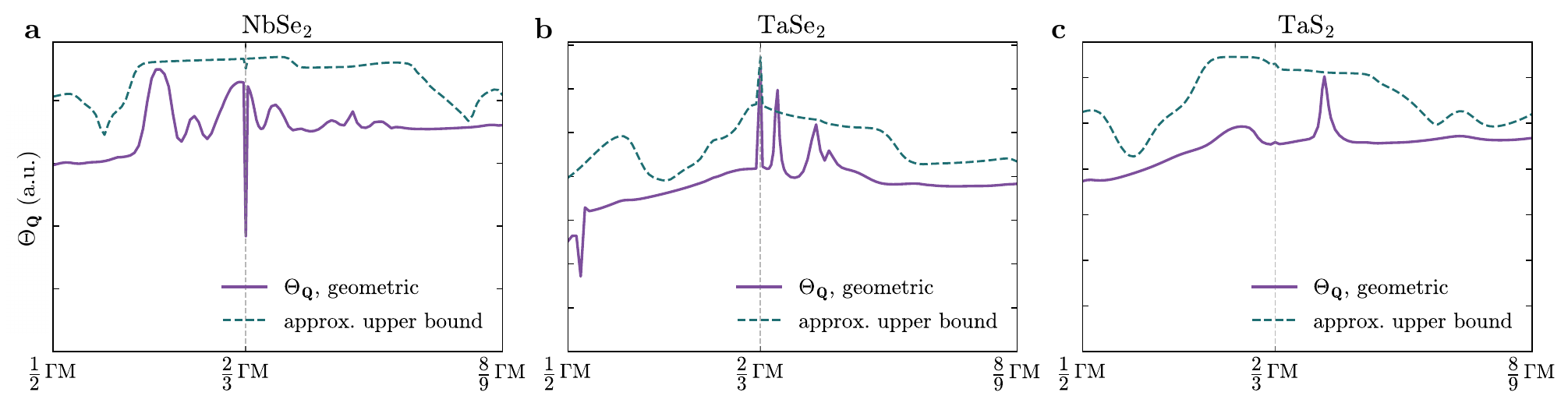}
    \caption{CDW driver $\Theta_\Q$ for \textbf{a} NbSe$_2$ ($T\approx 23 \, \text{K}$), \textbf{b} TaSe$_2$ ($T\approx 35 \, \text{K}$), and \textbf{c} TaS$_2$ ($T\approx 49 \, \text{K}$), overlaid with approximate upper bound derived upon inclusion of a sextic term in the GL functional. Major corrections are only needed at $2/3 \, \G \text{M}$ for NbSe$_2$ and near $1/2 \, \G \text{M}$ for TaSe$_2$, where the original driver exhibits strongly negative peaks from a negative $\beta_\Q$. The ordering wavevector estimates from the unmodified driver as specified in the main text are still unlikely to change. Additionally, the secondary peak above $2/3 \, \G \text{M}$ for TaSe$_2$ overshoots the upper bound due to small $\beta_\Q$ and is expected to become shorter.}
    \label{fig:overlay}
\end{figure}

Fig. \ref{fig:overlay} plots the driver $\Theta_\Q$ from the main text overlaid with the approximate upper bound derived here. While the strongly negative peaks in $\Theta_\Q$ at $2/3 \, \G \text{M}$ for NbSe$_2$ and near $1/2 \, \G \text{M}$ for TaSe$_2$ need to be corrected using the sextic term, the conclusions drawn in the main text regarding charge ordering wavevectors still stand. The only other correction is to the secondary peak in the TaSe$_2$ driver, which upon inclusion of a sextic term is likely to become significantly shorter.   

%\newpage

\section*{Supplementary Note 4: Anisotropy of the quadratic susceptibility $\chi$}

\begin{figure}[h]
    \centering
    \includegraphics[width=\linewidth]{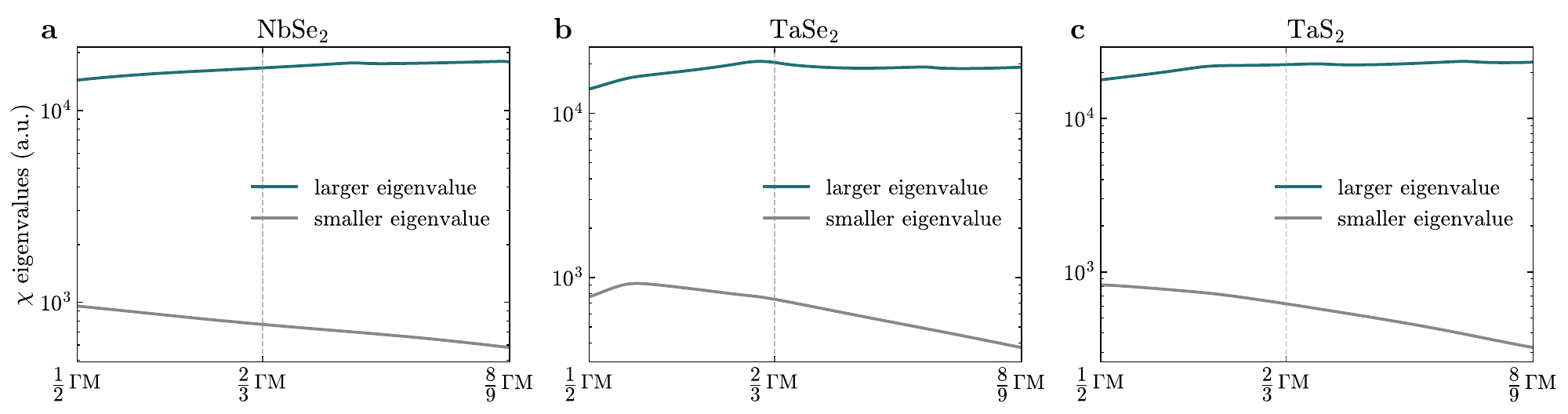}
    \caption{Both eigenvalues of the modified tensorial Lindhard susceptibility $\chi$ as a function of $\Q$ for \textbf{a} NbSe$_2$ ($T\approx 23 \, \text{K}$), \textbf{b} TaSe$_2$ ($T\approx 35 \, \text{K}$), and \textbf{c} TaS$_2$ ($T\approx 49 \, \text{K}$). Note that the eigenvalues have been plotted on a log scale, and the anisotropic nature of $\chi$ is evident for all three TMDs considered in this work.}
    \label{fig:chi_eigs}
\end{figure}

We explicitly find that the quadratic susceptibility is indeed highly anisotropic, that is, one of the eigenvalues of $\chi$ is much smaller than the other. It is then sufficient to work within the space of the larger eigenvalue (where the quadratic term in the GL free energy becomes close to zero near the transition temperature). In Fig. \ref{fig:chi_eigs}, we plot the eigenvalues of $\chi$, the symmetric rank-2 tensor that forms part of the coefficient of $\hat{\bm n}^{\otimes2} \Delta^2$ in the GL free energy (along with the interaction vertex $g_\Q$), and find that they indeed differ by at least one order of magnitude, thereby allowing the reduction of the free energy into a scalar order parameter $\Delta$ at leading order. 

%\newpage

\section*{Supplementary Note 5: Variation of the CDW driver $\Theta_\Q$ with temperature}
\begin{figure}[h]
    \centering
    \includegraphics[width=\linewidth]{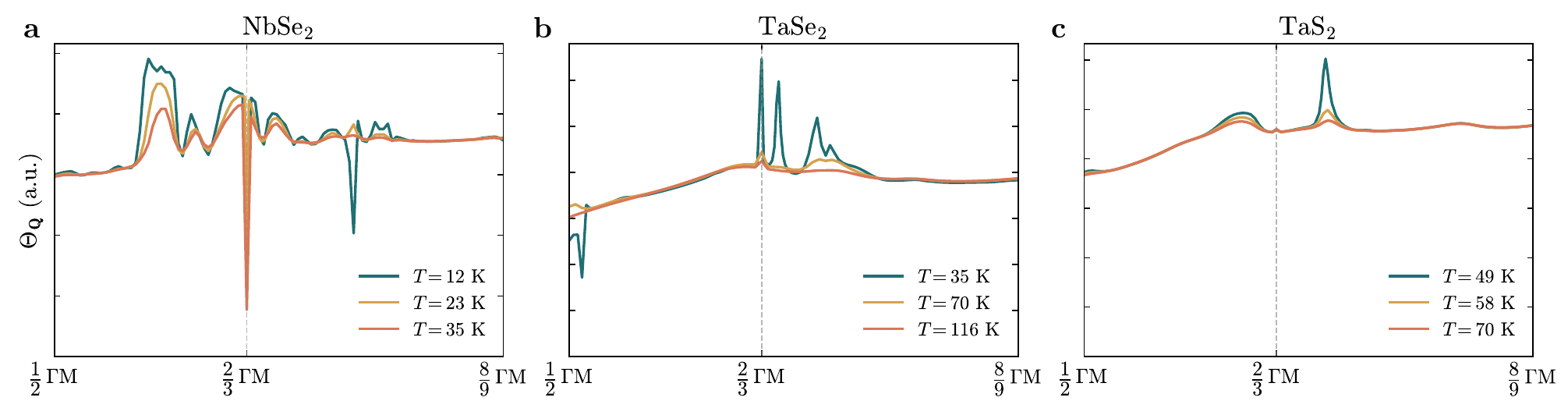}
    \caption{CDW transition driver $\Theta_\Q$ plotted as a function of $\Q$ for \textbf{a} NbSe$_2$, \textbf{b} TaSe$_2$, and \textbf{c} TaS$_2$ at different temperatures $T$. For all three TMDs, increasing $T$ smoothens out the peaks. However, the quadratic background (from $-\Pi_\Q$) remains largely constant with $T$, further proving that these geometry-enhanced higher-order-terms drive the CDW transition in these quasi-flatband hexagonal materials.}
    \label{fig:temp}
\end{figure}

Here, we present the variation of the CDW driver $\Theta_\Q$ with temperature (see Fig. \ref{fig:temp}). Recalling that the transition occurs when $3/\text{Tr}\, g_\Q = \Theta_\Q$ and that $g_\Q$ is independent of temperature, the variation of the driver with temperature signals when the charge ordering takes place. While the exact transition temperature cannot be estimated within the scope of this work since it depends on the phononic spectrum through $g_\Q$ (and is likely to get renormalized anyway due to higher order terms and other interactions), the trend does allow for qualitative conclusions to be made.

Firstly, as expected, higher temperatures smoothen out the peaks. Interestingly however, only the higher-order-terms in the driver seem to show any significant variation with temperature while the quadratic background largely stays the same. This is notable since it suggests that the transition is driven precisely by these same higher-order-terms; $\Pi_\Q$ being largely constant with temperature implies that it cannot possibly equal $3/\text{Tr}\, g_\Q$ by lowering the temperature, as would be required by a second-order transition driven solely by the quadratic susceptibility. Secondly, it is commonly reported in experiment \cite{SM:Moncton_nbse2_1975,SM:moncton_nbse2_1977,SM:FriendJerome1979pldcdw,SM:Liu_tas2_expt} that the transition is either second-order or weakly first order. Since the driver approaches the flat quadratic form at high temperatures, our theory is consistent with a weakly first-order transition; then, it is probable that $3/\text{Tr}\, g_\Q$ rests slightly above this quadratic background, and geometry-enhanced higher order terms raise the driver to the ordering threshold upon lowering temperature.

\end{document}